\documentclass[twocolumn,trackchanges]{aastex631}
\usepackage{lipsum}% for placeholder text
\usepackage{graphicx} % Required for \includegraphics
\usepackage{amssymb}
\usepackage{float}
\usepackage{placeins}

\newcommand\lsim{\mathrel{\rlap{\lower4pt\hbox{\hskip1pt$\sim$}}
\raise1pt\hbox{$<$}}}
\newcommand\gsim{\mathrel{\rlap{\lower4pt\hbox{\hskip1pt$\sim$}}
\raise1pt\hbox{$>$}}}

\begin{document}
\title{\Large Cataclysmic Variables in Triples: Formation Models and New Discoveries}

\author[0000-0003-1247-9349]{Cheyanne Shariat}
\affiliation{Department of Astronomy, California Institute of Technology, 1200 East California Boulevard, Pasadena, CA 91125, USA}
\affiliation{Department of Physics and Astronomy, University of California, Los Angeles, Los Angeles, CA 90095, USA}
\affiliation{Mani L. Bhaumik Institute for Theoretical Physics, University of California, Los Angeles, Los Angeles, CA 90095, USA }

\author[0000-0002-6871-1752]{Kareem El-Badry}
\affiliation{Department of Astronomy, California Institute of Technology, 1200 East California Boulevard, Pasadena, CA 91125, USA}

\author[0000-0002-9802-9279]{Smadar Naoz}
\affiliation{Department of Physics and Astronomy, University of California, Los Angeles, Los Angeles, CA 90095, USA}
\affiliation{Mani L. Bhaumik Institute for Theoretical Physics, University of California, Los Angeles, Los Angeles, CA 90095, USA }

\author[0000-0003-4189-9668]{Antonio C. Rodriguez}
\affiliation{Department of Astronomy, California Institute of Technology, 1200 East California Boulevard, Pasadena, CA 91125, USA}

\author[0000-0002-2626-2872]{Jan van Roestel}
\affiliation{Anton Pannekoek Institute for Astronomy, University of Amsterdam, 1090 GE Amsterdam, The Netherlands}


\correspondingauthor{Cheyanne Shariat}
\email{cshariat@caltech.edu}

\begin{abstract}
The formation of cataclysmic variables (CVs) has long been modeled as a product of common envelope evolution (CEE) in isolated binaries. However, a significant fraction of intermediate-mass stars -- the progenitors of the white dwarfs (WDs) in CVs -- are in triples. We therefore investigate the importance of triple star dynamics in CV formation. Using {\it Gaia} astrometry and existing CV catalogs, we construct a sample of $\sim50$ CVs in hierarchical triples within 1 kpc of the Sun, containing main-sequence (MS) and WD tertiaries at separations of $100 - 30,000$ au. We infer that at least $10\%$ of CVs host wide tertiaries. To interpret this discovery, we evolve a population of 2000 triples using detailed three-body simulations, 47 of which become CVs. We predict that $20\%$ of CVs in triples form without ever experiencing CEE, where the WD and donor are brought together by the eccentric Kozai-Lidov (EKL) mechanism after the formation of the WD. These systems favor larger donor stars and longer birth orbital periods ($8-20$~hrs) than typical CVs. Among systems that do undergo CEE, about half would not have interacted without the presence of the tertiary. Triple formation channels both with and without CEE require initially wide inner orbits ($\gtrsim 1$ au), which in turn require larger tertiary separations to be stable. Consistent with this prediction, we find that the observed {\it Gaia} CV triples have wider separations on average than normal wide binaries selected in the same way. Our work underscores the importance of triples in shaping interacting binary populations including CVs, ultracompact binaries, and low-mass X-ray binaries.
\end{abstract}

\section{Introduction}\label{sec:introduction}
Cataclysmic Variables (CVs) are interacting binary systems in which a white dwarf accretes from a main-sequence or partially evolved companion star. Their orbital periods span from approximately $\sim80$~min up to $1$~day, with a majority of systems found at periods of a few hours. \cite[e.g.,][]{Patterson84, Warner95}. 
To date, all evolutionary models of CVs (outside of dense clusters) have assumed that CVs form via common envelope (CE) evolution \citep[e.g.,][]{Paczynski76}.
In this framework, the primary star in the initially wide ($a\sim 1$~au) MS+MS progenitor binary evolves into a red giant (RG) and engulfs its companion. If the donor successfully ejects the envelope, the stars can emerge as a tight WD+MS binary. Subsequently, the MS star fills its Roche lobe, either due to angular momentum losses from magnetic braking and gravitational waves or as a result of its nuclear evolution, thereby forming the CV.

% At this stage, orbital angular momentum is lost through magnetic braking, among other loss mechanisms, which shrink the orbit and eventually initiate stable mass transfer from the MS to the WD, thereby forming the CV. 

The progenitors of most WDs in CVs have masses of a few M$_\odot$ \citep[e.g.,][]{Littlefair08, Zorotovic20}. A notable fraction \citep[$10-20\%$;][]{Evans05, Raghavan2010, Tokovinin14, Moe17, Offner23} of such stars are actually triples, and this fraction can be even higher ($\sim30\%$) considering that some of the wide third bodies have since become unbound \citep[e.g.,][]{Shariat23}.  Triple dynamics is known to significantly influence the evolution of other WD and MS binary populations \citep[e.g.,][]{Naoz2014, Toonen2016, Toonen17, Toonen20, Shariat23, Shariat24}. It thus stands to reason that triple dynamics may play an important role in the formation of CVs.
Yet, triples have been largely ignored in evolutionary models for CVs \citep[with the notable exception of a triple model for the recurrent Nova T Pyx;][]{Knigge22}, and there has never been a systematic observational study of CVs in triples. 
% Naturally, the CV population is also expected to have imprints of triple dynamics be shaped by triple dynamics. However, this aspect of CV evolution has received limited attention in the literature \citep{Pala20, Knigge22}.

In hierarchical triples -- where two stars are in a relatively close orbit (the inner binary) compared to the distant tertiary's orbit around the inner binary -- new dynamical phenomena are introduced, which alters the evolution of the inner binary. One major phenomenon in these systems is the eccentric Kozai Lidov (EKL) mechanism \citep[][]{Kozai1962,Lidov1962,Naoz2016}, through which the tertiary can drive cyclic variations in the eccentricity and inclination of the inner binary in a secular fashion. For example, EKL-induced oscillations from a faraway tertiary can induce extreme eccentricities to the inner binary, leading to small ($\sim1-10$~R$_\odot$) periastron separations. During these close passages, tides and mass transfer can shrink wide inner binaries and bring them into contact. This pathway can be a significant formation channel for many binary phenomena, such as black hole low-mass X-ray binaries \citep[BH-LMXBs;][]{NaozLMXB, Shariat24c}, WD collisions \citep[e.g.,][]{Thompson2011}, nearly all close MS binaries \citep[][]{Tokovinin06, Fabrycky07}, and several forms of WD or MS merger products \citep[][]{Lombardi02, Naoz2014, Cheng19, Toonen20, Heintz22, Shariat23, Shariat24}. 

% Post Common Envelope Binaries \citep[e.g.,][]{Toonen13, Zorotovic14, Yamaguchi24}.
% compact-object gravitational wave events \citep[][]{Xuan21,Xuan23,Xuan24 Shariat23, many other papers about triples+GWs}

In this paper, we construct a high-confidence sample of $\sim50$ CV triples and use it to explore CV formation in a three-body environment. First, in Section \ref{sec:observed_sample}, we discuss the sample of CV triples that we construct using {\it Gaia} DR3. Next, we run a suite of detailed three-body stellar evolution simulations and investigate CV formation in triples (Section \ref{sec:simulated_sample}). We then compare our theoretical simulations to observations in Section \ref{sec:results}. In Section \ref{sec:discussion} we discuss our results and their caveats. Lastly, in Section \ref{sec:conclusions} we summarize our main conclusions.

\begin{figure*}[ht]
    \centering
    \begin{minipage}[t]{0.49\textwidth}
        \includegraphics[width=\textwidth]{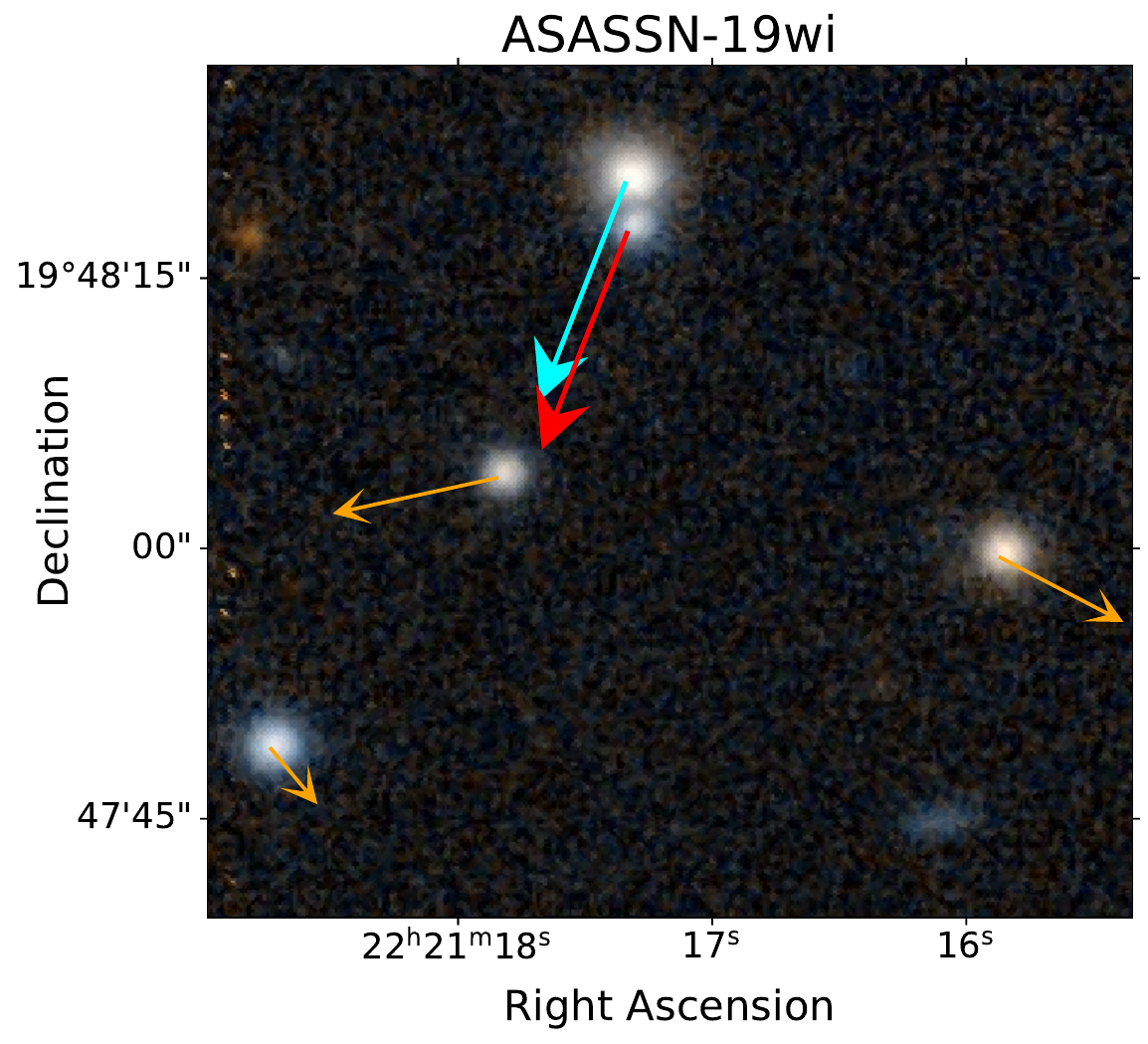}
    \end{minipage}
    \hfill
    \begin{minipage}[t]{0.49\textwidth}
        \includegraphics[width=\textwidth]{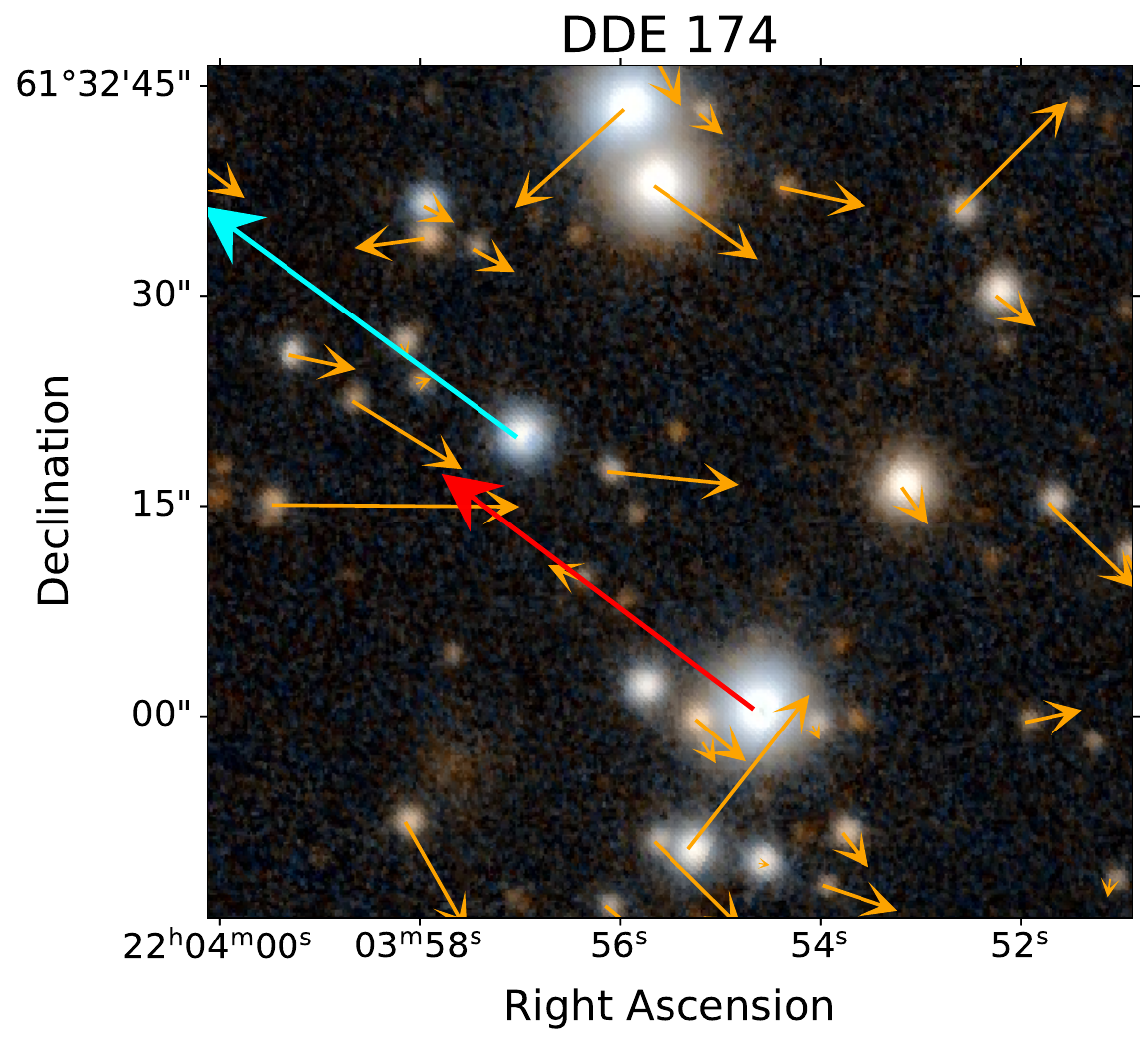}
    \end{minipage}
    \vspace{0.1cm}  % Adjust spacing between rows
    \begin{minipage}[t]{0.49\textwidth}
        \includegraphics[width=\textwidth]{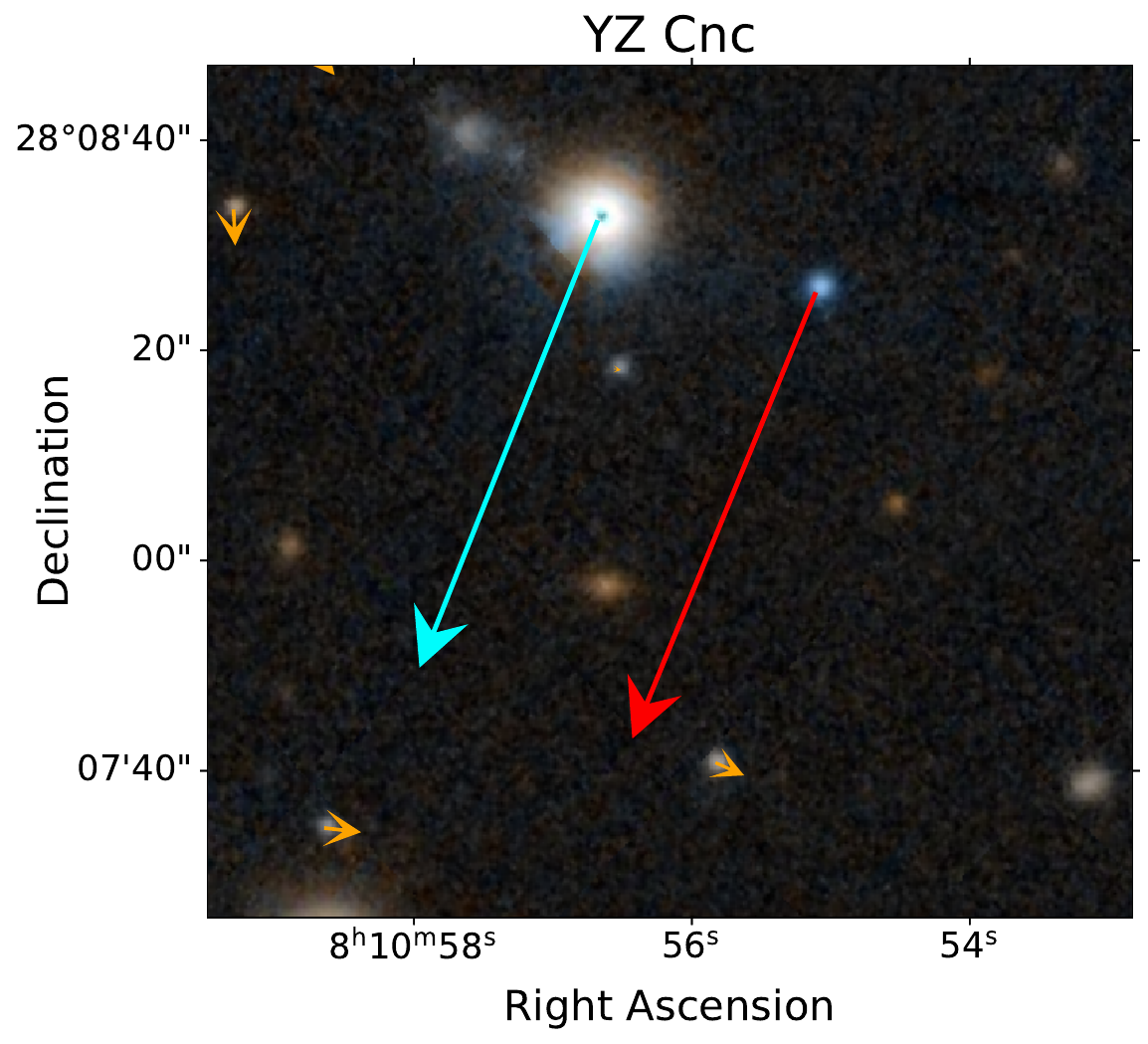}
    \end{minipage}
    \hfill
    \begin{minipage}[t]{0.49\textwidth}
        \includegraphics[width=\textwidth]{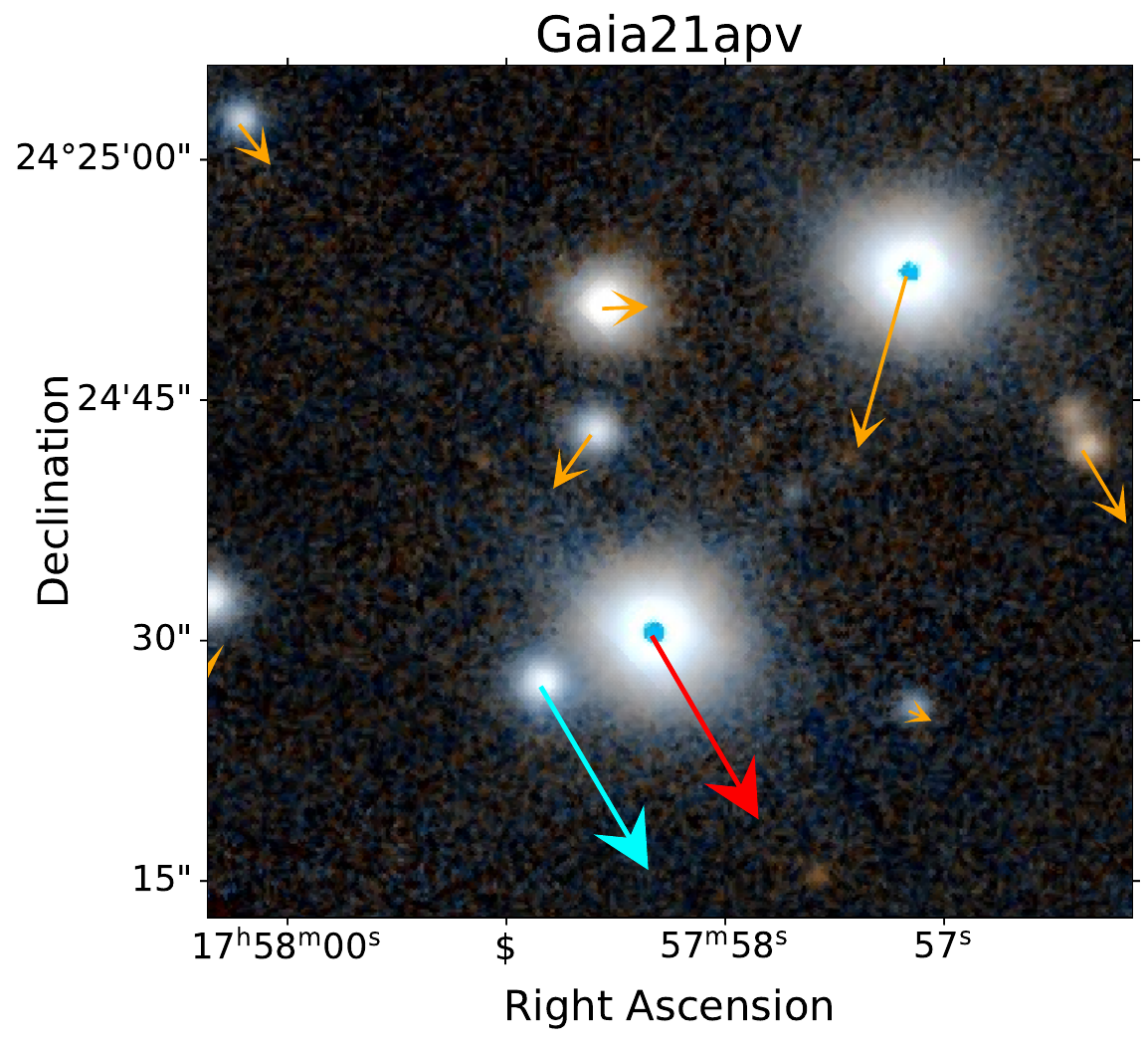}
    \end{minipage}
    \caption{Pan-STARRS1 images of CV triples with {\it Gaia} proper motion arrows indicated for the CV (blue) and the wide companion (red), along with other nearby sources (orange). The arrows are scaled arbitrarily for visual purposes. In this subset of our CV triple sample, we include ASASSN-19wi, DDE 174, YZ Cnc, and Gaia21apv. YZ Cnc (bottom left) has a white dwarf tertiary while the other three have main-sequence companions at a range of angular separations and flux ratios.}
    \label{fig:proper_motion_image}
\end{figure*}

\section{Observational Sample of CV triples}\label{sec:observed_sample}

\subsection{Constructing the Catalog}\label{subsec:catalog_construction}
To identify a sample of cataclysmic variables (CVs) in triple systems, we cross-match three partially overlapping CV catalogs with a {\it Gaia} eDR3 wide binaries catalog. The first is the Ritter and Kolb (RK, 7th edition, last updated 2015) catalog, which includes $1429$ CVs identified primarily through optical outbursts and X-ray associations \citep{Ritter03}. Unlike the following two catalogs, most of the CVs in RK have been followed up extensively and spectroscopically confirmed, leading to a high purity rate. The second source is the International Variable Star Index (VSX), from which we extract objects with CV classifications. Note that these classifications are often based on light curves and that some fractions are expected to be misclassified. The third catalog comprises CVs identified during outbursts via Zwicky Transient Facility (ZTF) transient alerts \citep[][van Roestel et al. in prep]{Szkody20}. After consolidating these three catalogs and removing duplicates, we cross-match this initial sample with {\it Gaia} wide binaries.
We search for wide stellar companions to the CVs by positionally cross-matching them to the wide binary catalog from \citet{EB21}. This catalog contains over a million spatially resolved binaries, where the two stars share a common proper motion and parallax in {\it Gaia} eDR3. The binary catalog is restricted to sources within $\approx1$~kpc of the Sun, so we only consider CVs with 1 kpc. 
Previous works \citep{Michaely2014,Igoshev20} proposed and carried out searches for common proper motion companions around post-CE binaries to constrain common-envelope evolution. Here, we adopt a similar methodology to study formation pathways of CVs, using the more precise {\it Gaia} DR3 astrometry and incorporating explicit chance alignment probabilities to identify wide companions with high confidence.

\citet{EB21} crudely classified the binaries in the catalog as `MSMS,' `WDMS,' and `WDWD' based on the components' position in the {\it Gaia} color-magnitude diagram, although there is expected contamination by unresolved inner binaries \citep[e.g.,][]{EB21, Nagarajan24Triples}.
Beyond MSMS binaries, which are the vast majority of the sample, the catalog also includes $15,982$ WDMS binaries, $1362$ WDWD binaries, and $\sim20,000$ binaries containing giants and subgiants. Most of the CVs in our sample fall between the MS and WD cooling track, with some classified as MS stars and others classified as WDs. The binary separations of the entire wide binary catalog range from a few au to around $1$~pc, with a peak near $1000$~au that is mainly a consequence of {\it Gaia}'s $\sim1$~arcsec angular resolution.

To perform the positional cross-match, we identify the CVs in our list that are within 1 arcsecond of one of the stars in the resolved binary. We then attribute the closest spatial component in the binary to being the CV. We also manually check this scheme, finding that three sources were matched to the wrong component in the binary, which we correct. After cross-matching our combined CV catalog to the {\it Gaia} wide binary catalog, we only keep the binaries with {\tt R\_chance\_align} $<0.1$, namely less than a $10\%$ chance alignment probability \citep[e.g.,][]{EB18,EB21}. Most of the systems in our sample have {\tt R\_chance\_align} significantly smaller than this $10\%$ limit, with a median value of $2\times10^{-4}$. Our final sample contains $49$ CVs with resolved wide tertiary stellar companions from {\it Gaia}: `CV triples.' Figure \ref{fig:proper_motion_image} illustrates our method of identifying gravitationally bound companions to CVs. Table \ref{tab:catalog_CVs} summarizes the parent samples used in our CV search, including the number of CVs within 1 kpc (our search volume), those with resolved companions, and the triple fraction. Note that this triple fraction is heavily underestimated due to selection biases, as discussed in Sections \ref{subsec:selection_biases} and \ref{subsec:CV_Triple_Fraction}.

\begin{deluxetable}{lccc}[hb]
\tablecaption{Parent Samples of CVs\label{tab:catalog_CVs}}
\tablehead{
\colhead{Catalog} &
\colhead{{\it Gaia}} &
\colhead{N} &
\colhead{Triple}
\\
\colhead{} &
\colhead{($< 1$ kpc)} &
\colhead{Triple} &  
\colhead{Fraction} 
}
\startdata
RK  & 841 & 27 & 0.032   \\
VSX & 2437 & 45 & 0.018  \\
ZTF & 2172 & 34 & 0.016  
\enddata 
\end{deluxetable}
% \vspace{-3em} % Reduce vertical space

In Table \ref{tab:CV_comps_mini}, we summarize the basic properties of our CV triples sample. Properties include the literature name of the source, {\it Gaia} source IDs for both the CV and companion, the CV's parallax ($\varpi_{\rm CV}$), the projected separation between the CV and companion ($s$), and the parent catalogs that they are included in. For a full machine-readable version of this table -- which also includes absolute {\it Gaia} G magnitude of the CV, the BP-RP color of the CV, the period reported by VSX, and the period that we derived from ZTF -- please refer to in Appendix \ref{app:data} and Table \ref{tab:CV_gaia_params_app}. 

The CV triples have separations ranging from $150$~au to $10^4$~au, with a median separation of $2765$~au. These separations are limited by {\it Gaia}'s angular resolution, which sets the minimum observable separation for a companion at a given distance. For example, in Figure \ref{fig:proper_motion_image}, we show some resolved CV triples, showing that those with smaller angular solutions or fainter components are missed in our sample. We discuss this selection bias and its effect on our sample in the following section.

\startlongtable
\begin{deluxetable*}{lcccccc}
\movetableright=-0.5in
    \tablecaption{CVs with {\it Gaia} Wide Companions\label{tab:CV_comps_mini}.}
    \tablehead{
    \colhead{Literature Name} &
    \colhead{CV} &
    \colhead{$\varpi_{\rm CV}$} & 
    \colhead{$s$} &
    \colhead{Catalog} &
    \colhead{Spectroscopically} &
    \colhead{Period} \\
    \colhead{} &
    \colhead{{\it Gaia }DR3 ID} &
    \colhead{[mas]} &  
    \colhead{[au]} &  
    \colhead{}   &
    \colhead{Confirmed?} &
    \colhead{[hr]}
    }
    \startdata
{\it Gaia} J051903.9+630339.6 &285957277597658240 &$8.585 \pm 0.025$ &$793.1$ &VSX, ZTF &(1) &-- \\
V3885 Sgr &6688624794231054976 &$7.753 \pm 0.034$ &$890.6$ &RK, VSX &(2) &4.97184 \\
V379 Tel &6658737220627065984 &$7.714 \pm 0.057$ &$4083.1$ &RK, VSX, ZTF &(1, 3) &-- \\
V1108 Her &4538504384210935424 &$6.475 \pm 0.135$ &$155.7$ &RK, VSX, ZTF &(4) &-- \\
MASTER OT J042609.34+354144.8 &176429285061830144 &$5.375 \pm 0.050$ &$3089.8$ &RK, VSX, ZTF &(5) &-- \\
SDSS J101421.55+063857.7 &3873721404734233344 &$5.281 \pm 0.193$ &$1654.7$ &ZTF &(6) &-- \\
LS IV -08 3 &4339398736975240192 &$4.673 \pm 0.044$ &$989.5$ &RK, VSX &(7) &4.68695 \\
ATO J145.8742+09.1629 &3855044176807504768 &$4.330 \pm 0.018$ &$1178.8$ &VSX &-- &19.46712 \\
YZ Cnc &683908812437792000 &$4.245 \pm 0.033$ &$5149.8$ &RK, VSX, ZTF &(8) &-- \\
USNO-A2.0 0825-10062737 &4364235192818015744 &$4.125 \pm 0.259$ &$317.2$ &VSX, ZTF &-- &1.272\tablenotemark{*} \\
KT Per &405873692214746368 &$4.059 \pm 0.034$ &$3838.0$ &RK, VSX &(9) &3.90379 \\
EF Peg &1759321791033449472 &$3.462 \pm 0.233$ &$1600.4$ &RK, VSX, ZTF &(10) &-- \\
UW Pic &4798833587650467200 &$3.422 \pm 0.027$ &$1502.3$ &RK, VSX &(11) &2.22336 \\
CSS 151208:020401+434133 &346478795636689024 &$3.358 \pm 0.104$ &$3756.7$ &VSX, ZTF &-- &1.152\tablenotemark{*} \\
SDSS J120615.73+510047.0 &1548430890981046272 &$3.299 \pm 0.165$ &$2034.3$ &RK &(12) &3.792\tablenotemark{*} \\
ASASSN-19dn &5914778653652430464 &$3.266 \pm 0.078$ &$591.2$ &VSX, ZTF &-- &-- \\
RX J0131.4+3602 &319016667370280064 &$3.263 \pm 0.305$ &$8663.0$ &ZTF &-- &-- \\
ASASSN-21cw &4290033344968345088 &$3.192 \pm 0.042$ &$703.6$ &VSX, ZTF &-- &8.544\tablenotemark{*} \\
RX J0154.0-5947 &4714563374364671872 &$3.076 \pm 0.027$ &$2728.3$ &RK, VSX &(13) &1.33440 \\
MR UMa &772038105376131456 &$3.037 \pm 0.107$ &$1995.7$ &RK, VSX, ZTF &(14) &-- \\
IGR J19308+0530 &4294249387962232576 &$3.013 \pm 0.027$ &$722.1$ &VSX &-- &14.66208 \\
{\it Gaia} DR3 2802565013608348928 &2802565013608348928 &$2.406 \pm 0.018$ &$519.5$ &ZTF &-- &7.776\tablenotemark{*} \\
DDE 202 &4472788803304480896 &$2.404 \pm 0.063$ &$2573.3$ &VSX &-- &4.056\tablenotemark{*} \\
MASTER OT J072948.66+593824.4 &1085880432371109888 &$2.361 \pm 0.269$ &$4231.4$ &RK, VSX, ZTF &-- &4.2\tablenotemark{*} \\
DDE 174 &2203373473312011008 &$2.329 \pm 0.083$ &$11370.3$ &ZTF &-- &-- \\
HW Boo &3741979368899159552 &$2.291 \pm 0.135$ &$26296.4$ &RK, VSX, ZTF &(14) &-- \\
DDE 169 &377904796464796416 &$2.268 \pm 0.097$ &$12985.3$ &VSX, ZTF &-- &-- \\
ASASSN-14hw &5549916268416336512 &$2.253 \pm 0.117$ &$25469.4$ &VSX, ZTF &-- &-- \\
DDE 202 &4580881692647497088 &$2.174 \pm 0.057$ &$3548.2$ &VSX, ZTF &-- &3.552\tablenotemark{*} \\
J2256+5954\tablenotemark{a} &2014349389931360768 &$2.065 \pm 0.017$ &$2825.6$ &RK &(15) &5.48654 \\
AH Men &5207384891323130368 &$2.043 \pm 0.014$ &$1439.1$ &RK, VSX &(16) &3.05304 \\
HS 0218+3229 &325051822271077376 &$2.018 \pm 0.042$ &$2801.9$ &RK, VSX, ZTF &(17) &-- \\
ZTF18abmneqb &2861094289593259136 &$1.946 \pm 0.173$ &$5775.1$ &VSX, ZTF &-- &1.776\tablenotemark{*} \\
ASASSN-19pz &5391459325544320640 &$1.938 \pm 0.234$ &$2520.6$ &VSX, ZTF &-- &-- \\
MASTER OT J084404.01+794408.7 &1144277041110541440 &$1.878 \pm 0.096$ &$12765.8$ &VSX, ZTF &-- &2.52\tablenotemark{*} \\
BP CrA &6733050836430023552 &$1.780 \pm 0.042$ &$27450.2$ &VSX &-- &-- \\
RZ Gru &6544371342567818496 &$1.777 \pm 0.020$ &$1049.0$ &RK, VSX &(18) &8.64000 \\
VZ Scl &2337436792938619392 &$1.735 \pm 0.041$ &$32577.8$ &RK, VSX &(19) &3.47093 \\
ASASSN-14eq &4918835764173746432 &$1.586 \pm 0.076$ &$27415.9$ &RK, VSX, ZTF &(20) &-- \\
NSV 7184 &5985406470331309312 &$1.461 \pm 0.043$ &$11513.1$ &VSX, ZTF &-- &-- \\
ASASSN-18ed &6139397368696634240 &$1.381 \pm 0.142$ &$9604.1$ &VSX, ZTF &-- &-- \\
AY Psc &2565601982736199168 &$1.368 \pm 0.043$ &$28801.9$ &RK, VSX, ZTF &(21) &5.232\tablenotemark{*} \\
SDSS J080033.86+192416.5 &670132550216853632 &$1.335 \pm 0.245$ &$3334.6$ &RK, VSX, ZTF &-- &-- \\
ASASSN-19wi &1777641510176281600 &$1.333 \pm 0.058$ &$2064.5$ &VSX, ZTF &-- &-- \\
ASASSN-14jd &5269753451459457024 &$1.257 \pm 0.110$ &$936.1$ &VSX, ZTF &-- &-- \\
ASASSN-19eu &6099000383784197120 &$1.176 \pm 0.090$ &$1404.2$ &VSX, ZTF &-- &-- \\
SWIFT J2124.6+0500 &2699191408560964736 &$1.136 \pm 0.022$ &$5279.2$ &VSX &(22) &19.992 \\
TX Col &4804695423438691200 &$1.020 \pm 0.020$ &$2525.0$ &RK, VSX &(23) &5.7192 \\
        \enddata
        \vspace{1mm}
        {\bf References:} (1) \citet{Pala20}; (2) \citet{Linnell09}; (3) \citet{Potter05}; (4) \citet{Price04}; (5) \citet{Han20}; (6) \citet{Inight23}; (7) \citet{Bruch17}; (8) \citet{Oke82}; (9) \citet{Clarke84}; (10) \citet{Howell94}; (11) \citet{Romero-Colmenero03}; (12) \citet{Schwope09}; (13) \citet{Beuermann21}; (14) \citet{Thorstensen12}; (15) \citet{Kjurkchieva15}; (16) \citet{Froning12}; (17) \citet{Rodriguez09}; (18) \citet{LaDous90}; (19) \citet{Odonoghue87}; (20) \citet{Paterson19}; (21) \citet{Szkody93}; (22) \citet{Halpern13}; (23) \citet{Buckley87}.
        \tablenotetext{a}{This CV resides in a quadruple star system, where its wide companion is an unresolved binary.}
        \tablenotetext{*}{Denotes the best-fit period from the ZTF optical light curve for sources without reported VSX periods.}
\end{deluxetable*}

\vspace{-4em} % Reduce vertical space
\subsection{Selection Biases}\label{subsec:selection_biases}

\begin{figure*}
\centering
\includegraphics[width=0.99
\textwidth]{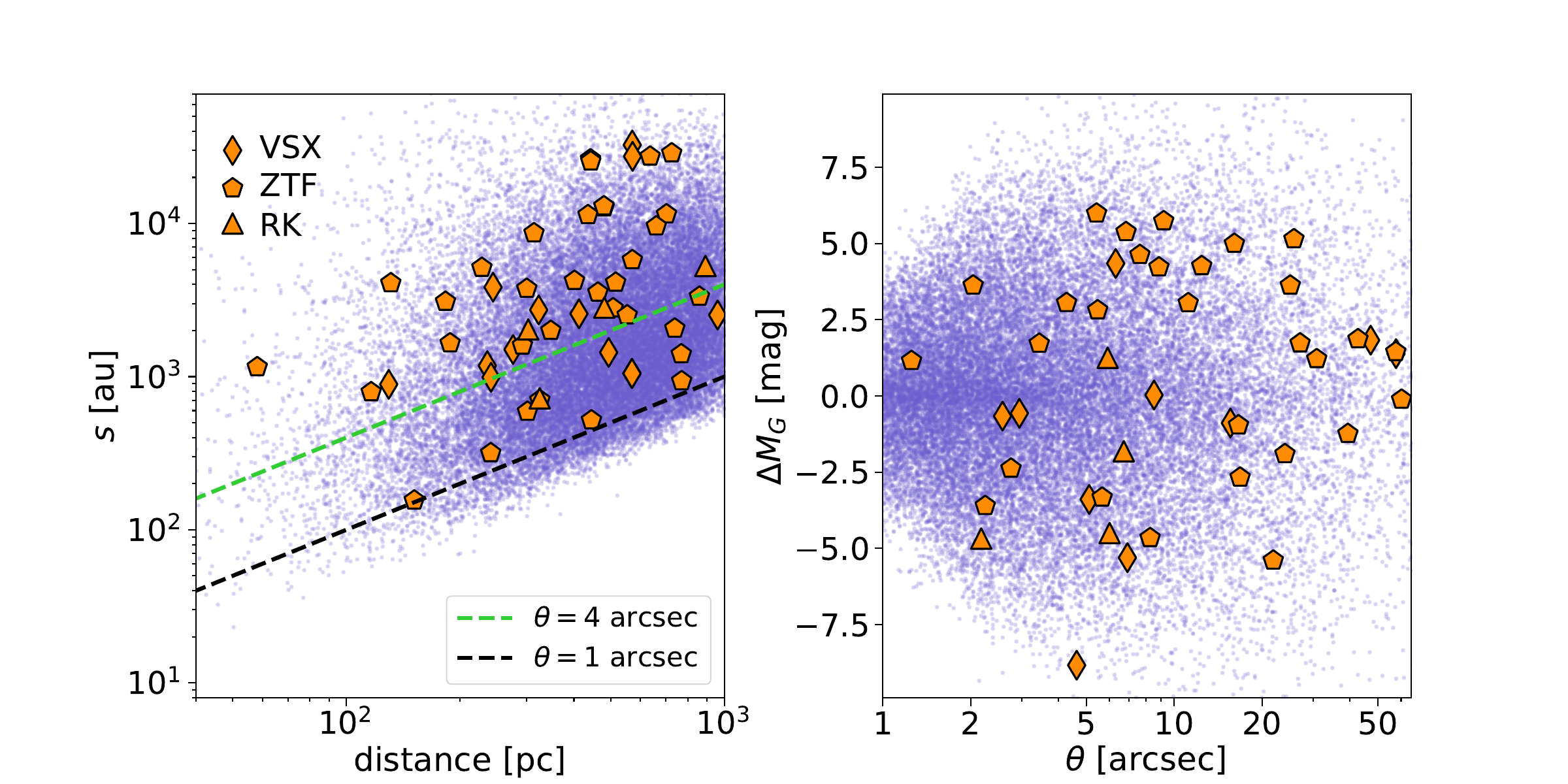}
\caption{Distance vs separation (left) and angular separation vs. difference in G magnitude (right) for all {\it Gaia} CVs. In purple, we plot a randomly selected sample of $10^5$ {\it Gaia} wide binaries from \citet{EB21}, chosen such that the chance alignment probability is less than $10\%$, as is the case with our CVs. In orange, we show the CV triples. In the left panel, the black (green) lines denote the 1" and 4" angular resolution cutoffs (denoted by $\theta$). On the right, we plot the G-band flux difference ($\Delta M_{\rm G}$) between the G color of the CV and its tertiary companion as a function of $\theta$ (in log-scale). The sign of $\Delta G$ is randomized for the comparison sample.
Note that CV triples tend to be wider than the corresponding wide binaries in the field. We analyze this trend further in Section \ref{subsec:sep_distribution}.
}\label{fig:distance_sep_CVs} 
\end{figure*}

Our method of identifying CV triples using wide binary companions introduces several selection biases, leading to the omission of a significant fraction of bound tertiary companions. The first bias is that the tertiary needs to be resolved, which constrains its angular separation and flux ratio. In Figure \ref{fig:distance_sep_CVs}, we plot the distance of the CV triple as a function of physical separation between the CV and its companion. The black (green) line denotes the angular resolution cutoff for $1''$ and $4''$, and the purple points are a randomly selected sample of wide binaries in the \citet{EB21} catalog. Most of the CV tertiaries have $\theta > 4$ arcsec, which is wide enough that their BP-RP colors are mostly uncontaminated and their astrometry is negligibly affected by the presence of the companion. From the right panel of Figure \ref{fig:distance_sep_CVs}, we find that CV triples typically have larger flux ratios than wide binaries, and in about half of the cases, the CV is fainter than its companion. 
The small fraction of similar-flux tertiaries might reflect a selection bias for CVs, whereas nearby companions with comparable brightness might make it more difficult to detect or identify a source as a CV.

The second major selection effect is that companions must satisfy the condition {\tt parallax\_over\_error} $>5$ and be detected by {\it Gaia}. The first requirement ensures well-constrained parallax measurements and was used in constructing the wide binary catalog \citep{EB21}. These selection criteria introduce a bias against faint WD or M-dwarf companions to CVs, as such companions are often too faint to be detected by {\it Gaia}, especially at larger distances. Even when detected, their measurements are more likely to be less robust (i.e., larger errors), making them less likely to meet the 
{\tt parallax\_over\_error} $>5$ threshold. 

The third major selection effect is that tertiaries need to be confidently bound, leading to a bias against the widest companions ($s>20,000$~au). At these separations, it is difficult to statistically confirm the bound nature of a tertiary companion. As such, many of the bound tertiaries at wide separations likely have calculated {\tt R\_chance\_align}$>0.1$, leading us to exclude them from our sample of confidently bound CV triples. This bias affects the wide binaries sample as well, which we use for comparison later in the paper.

\begin{figure*}
    \centering
    \includegraphics[width=1.0
    \textwidth]{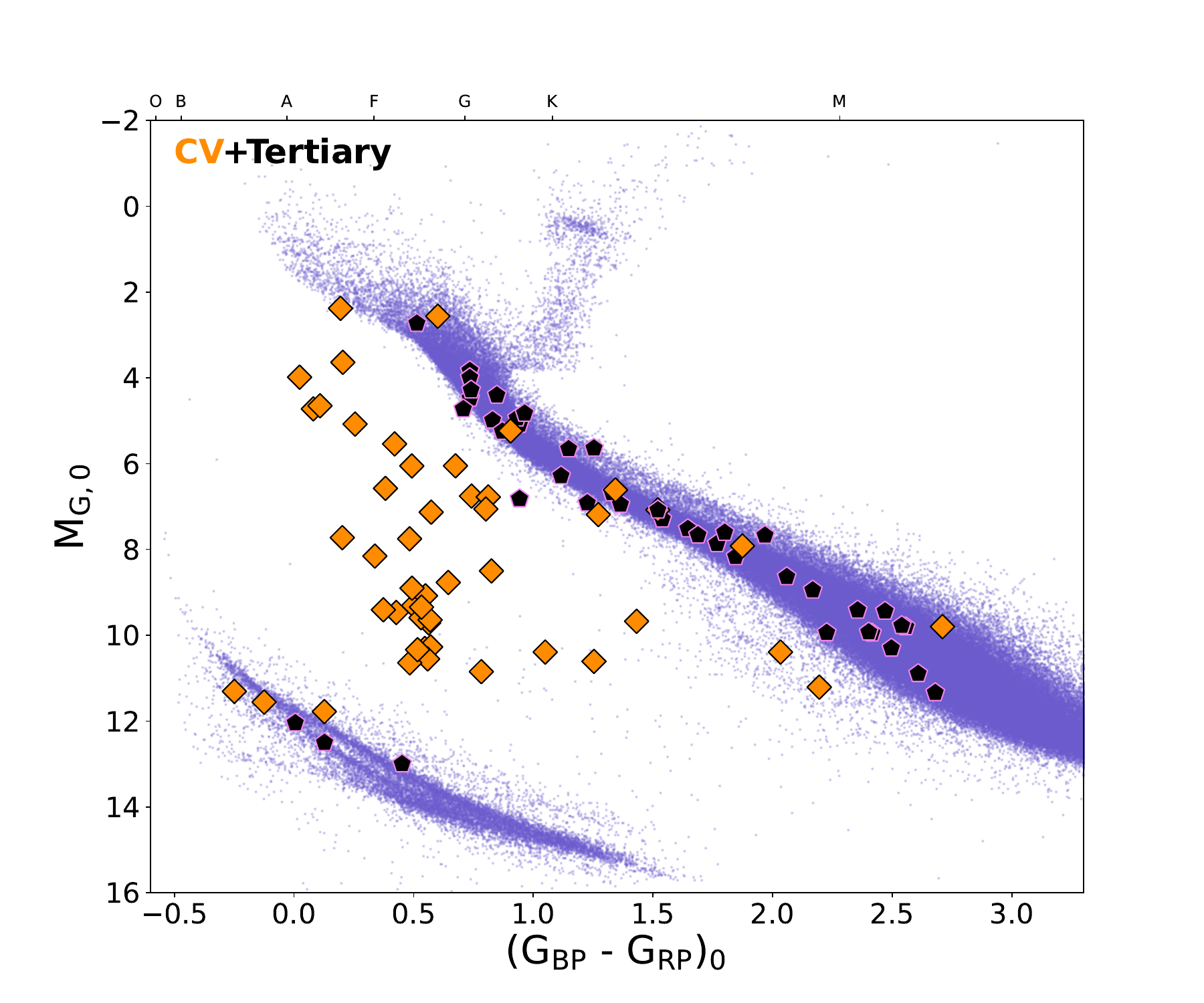}
    \caption{Extinction-corrected color-magnitude diagram of CVs (orange) and their wide tertiary companions (black). These CVs are identified from the VSX, ZTF, and RK catalogs and have resolved {\it Gaia} proper motion companions. The purple background points are {\it Gaia} sources within $100$~pc. On the top x-axis, we display the stellar types that correspond to different $G_{\rm BP}-G_{\rm RP}$ colors. The y-axis shows the absolute magnitude for the sources in the {\it Gaia} G band.
    }\label{fig:CV_HR_Diagram} 
\end{figure*}

\subsection{HR Diagram of CV triples}\label{subsec:Gaia_HR}
In Figure \ref{fig:CV_HR_Diagram} we plot an observational HR diagram of our CVs (orange) and their companions (black) using the {\it Gaia} colors. To correct for dust reddening, we use the 3D dust map from \citet{Edenhofer24}, who map the interstellar dust distribution out to $1.25$ kpc (a volume that includes all of the systems in our sample) using {\it Gaia }XP spectra. On the top axis of the HR diagram, we show the spectral types as derived from the {\it Gaia} BP-RP magnitude. Most of the tertiary companions are FGKM stars, and $3$ of the companions are white dwarfs. Since CVs are generally several Gyr old \citep[][]{Kolb96, Ak10, Canbay24}, main-sequence companions above a few solar masses are not expected since they would have already become WDs. 
% Note that this also introduces a bias to our tertiary detection, where faint white dwarf tertiaries are harder to detect at distances beyond a few hundred pc. 
The three CVs with WD tertiary companions are all relatively close, $240$, $230$, and $300$~pc away, and have tertiary separations of $1000$, $5200$, and $2100$~au, respectively. We, therefore, expect that many more CVs harbor wide WD companions that remain undetected. 

Most of the CVs in the sample lie between the main-sequence and the WD cooling track, with five lying on the main sequence and three on the WD track (Figure \ref{fig:CV_HR_Diagram}). 
% In the \citet{EB21} catalog, $39$ of the CVs triples were classified as `MSMS', $5$ are `WDMS', $5$ are `MS??', and $1$ is `WD??', based on their HR location.
Those on the main-sequence may correspond to CVs with somewhat evolved donors \citep{Abril20}. In this case, the donor dominates the measured flux in the optical bands. The overdensity of CVs with $M_{G,0}\sim10$, $(G_{BP}-G_{RP})_0\sim0.75$ correspond to classical dwarf novae \citep[e.g., Figure 2 of][]{Abril20}, where a combination of the accretion and stellar flux dominates in the optical. The CVs with $M_G < 6$ are `nova-like,' characterized by their high accretion rates where the accretion disk dominates the optical photometry. 

Interestingly, we find that one of the spectroscopically confirmed CVs, J2256+5954 (also known as UCAC4 750-078221), is part of a quadruple system. Its wide companion (GSC 03997-00231) is located $\sim2800$~au away and exhibits sinusoidal variability with three distinct periods: $14.49$, $20.79$, and $36.68$ hours. This companion lies above the main sequence (Table \ref{tab:CV_gaia_params_app} and Figure \ref{fig:CV_HR_Diagram}), and has been previously identified as a BY Dra variable \citep{Chen20}, suggesting it is consistent with an unresolved binary.

\subsection{ZTF Light Curves}\label{subsec:new_periods}
For each of our identified CVs, we search for optical time-domain photometry from the Zwicky Transient Facility \citep[ZTF;][]{ZTF19}. ZTF scans the northern sky (Declination$\gsim-28^\circ$) with $\sim2$~day cadence. We query the ZTF DR21 catalog using a basic cone search with a radius of 2'', since this is the median FWHM of the PSF of ZTF images. We find that $29$ of the CVs in our triples sample have ZTF light curves in the g and r bands (Figure \ref{fig:ztf_lcs} and Appendix \ref{app:data}). Over the six year ZTF baseline, most of the CVs exhibit some outbursting behavior and/or long-term variability. Several of them also show eclipses, with another fraction showing high/low state transitions during the $\sim5$ year of the ZTF observational baseline. This strong variability is expected for most classes of CVs. Two of the systems, {\it Gaia }DR3 2802565013608348928 and {\it Gaia }DR3 3855044176807504768 show no large outbursts or significant variability during the six year ZTF baseline, which may signify low mass transfer rates, or that these objects are not CVs.

For all the CVs with ZTF light curves, we perform a periodicity search. We first clip any outbursts by masking points that are two standard deviations above the mean magnitude. Then, we apply a Lomb Scargle (LS) periodogram and select the five periods with the largest LS powers in the 5 minute-24 hours period range, ignoring sidereal harmonics. We then phase-fold the light curves on these periods and manually identify whether any of the top five periods lead to smooth phase-folded light curves. We find some systems with robust periods, which we note in Table \ref{tab:CV_comps_mini}. We also plot a subset of this sample in Figure \ref{fig:phase_folded_LCs} and discuss their features in Appendix \ref{app:folded_LCs}. However, most of the CVs in our sample are significantly variable on a range of timescales, and we do not recover any modulations in our period range for these. This is not unusual for CVs, which often have aperiodic optical variability dominated by the accretion disk. For CVs that already contain a documented period in VSX, we generally find that the reported period is correct and matches the one we find or a factor of two discrepant. For four of our CVs, we find periodicities that were not previously reported in VSX and show their phase-folded light curves in Figure \ref{fig:phase_folded_LCs}. 

\subsection{Purity of the Sample}

The majority of the CVs in our sample appear in multiple catalogs, have light curves displaying clear outbursts, and fall between the main sequence and WD cooling track in the HR diagram (Figure \ref{fig:CV_HR_Diagram}), as expected for CVs. A handful of cases remain to be confirmed as CVs, and we elaborate on each of these below.

%Smadar is here

The three CVs that lie on the WD track are SDSS J101421.55+063857.7, RX J0131.4+3602, and {\it Gaia }DR3 4364235192818015744; the former two are spectroscopically confirmed CVs \citep{Jiang00, Inight23}, and SDSS J101421.55+063857.7 has a known orbital period of $85$ minutes \citep{Inight23}. On the other hand, {\it Gaia }DR3 4364235192818015744 has not been spectroscopically confirmed. The source is a WD candidate in the \citet{Gentile-Fusillo21} catalog and its ZTF light curve displays variability that may be consistent with a low-amplitude outburst  (Figure \ref{fig:ztf_lcs}). The CV nature of this system remains to be confirmed. % KE removed the 6 hour period bit because it doesn't make sense at this CMD position, and I don't see it convincingly in the light curve. 

%was noted previously as a white dwarf \citep{Gentile-Fusillo21} and we manually identify a $\sim6$ hour periodicity for this system with modest outbursts in the light curve (Figure \ref{fig:ztf_lcs}). The CV nature of this system remains to be confirmed.

Five of the CVs lie directly on the main-sequence track after extinction corrections. The first  is IGR J19308+0530, which was previously classified as an LMXB and has a bright F-star donor. This system is likely an accreting binary, and we discuss it further in Appendix \ref{app:IGR}. 
The second is {\it Gaia }DR3 2802565013608348928, which does not exhibit any major outburst in the ZTF light curve (Figure \ref{fig:ztf_lcs}). 
The third CV on the main-sequence is ATO J145.8742+09.1629 (or {\it Gaia }DR3 3855044176807504768), which is identified as `Nova-like' in VSX, and has previously been classified as a variable star \citep[][]{Heinze18}. This source also does not exhibit any major outbursts, and has a clear period of $9.74$ hours in the phase-folded ZTF light curve\footnote{There exists an eROSITA X-ray source a few arcseconds away with a soft X-ray flux of $9.3111\times10^{-14}$~erg~s$^{-1}$~cm$^{-2}$. Assuming this corresponds to the variable, which has {\it Gaia} parallax of $4.3$ mas, the has an X-ray luminosity of $5.94\times10^{29}$~erg~s$^{-1}$. This is consistent with a low accretion rate CV \citep[e.g.,][]{Rodriguez24} or an active binary.}. 
The last two sources with the faintest $G$-band absolute magnitudes on the main sequence are ASASSN-19dn and ASASSN-21cw, both of which are classified as U Geminorum-type variables in VSX. Both of these CV candidates have close, marginally resolvable companions ($590$ and $700$~au), and their companions do not have a reported BP-RP color. ASASSN-21cw does have a clear outburst in both the ZTF {\it r} and {\it g} bands, making it a strong CV candidate. ASASSN-19dn is outside the ZTF footprint but has a candidate outburst in its ASASSN light curve, where the {\it g} magnitude brightens from 16.4 to 15.8.
While these handful of sources remain to be confirmed, the vast majority of our CV sample remains as high-confidence CVs. 

\begin{figure}
    \centering
    \includegraphics[width=0.49
    \textwidth]{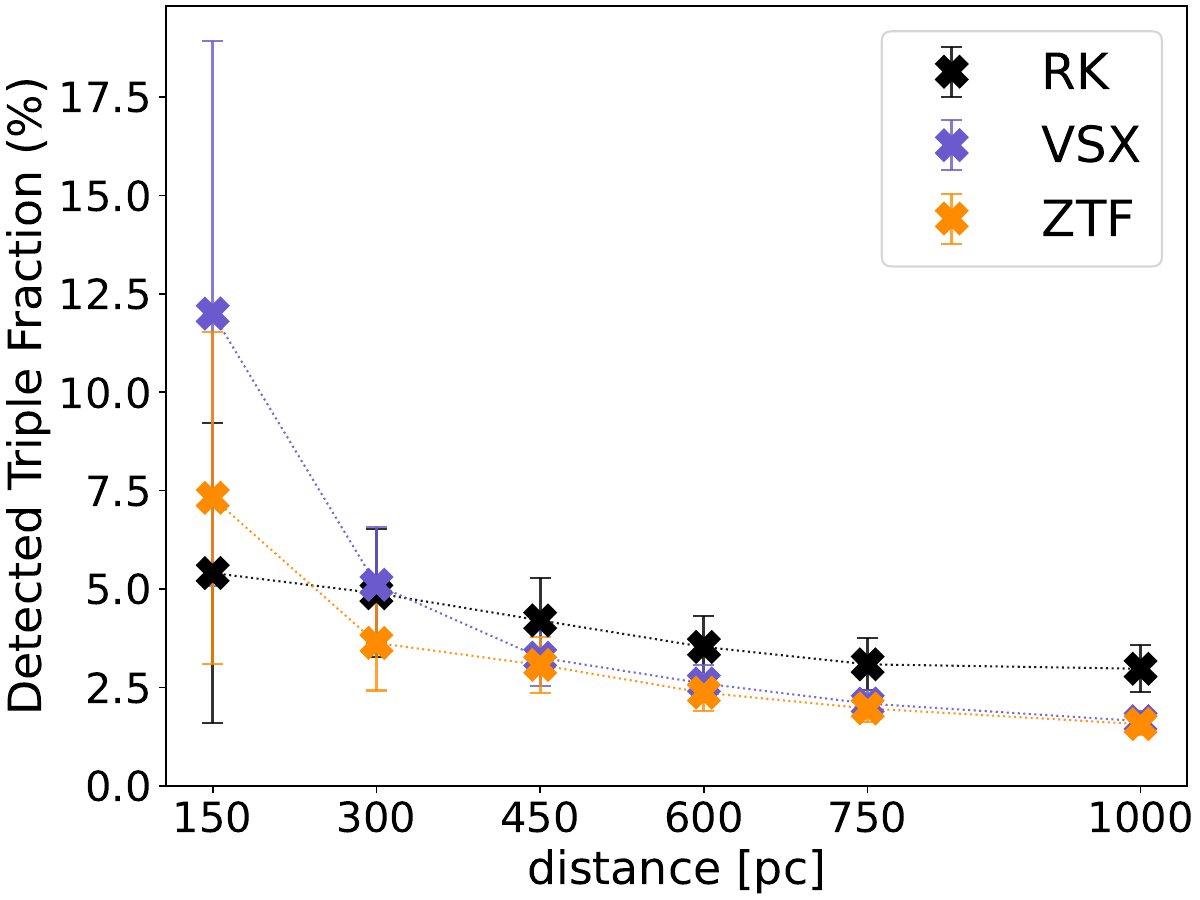}
    \caption{Fraction of CVs that reside in triples in various volume-limited sub-samples. For each bin, we plot the fraction of CVs within the given distance bin that host wide tertiary companions for classified CVs in the RK, VSX, and ZTF catalogs. Poisson errors are plotted for each bin. As the distance increases, resolving tertiary companions becomes increasingly difficult, decreasing the fraction of resolved outer companions.
    }\label{fig:CV_triple_fraction} 
\end{figure}

\subsection{CV Triple Fraction}\label{subsec:CV_Triple_Fraction}
As described in Section \ref{subsec:selection_biases}, several selection biases make it difficult to observe CV tertiaries, especially at larger distances. This makes it difficult to confidently estimate the true triple fraction of CVs. Nevertheless, by focusing on volume-limited sub-samples, we estimate lower limits for the triple fraction of CVs using our dataset.

In Figure \ref{fig:CV_triple_fraction}, we plot the CV triple fraction ($N_{\rm \text{CV, triple}} / N_{\rm \text{CVs}}$) for different volume limited sub-samples. We also plot Poisson errors for each bin ($\sqrt{N_{\rm \text{CV, triple}}} / N_{\rm \text{CVs}}$). At closer distances, selection effects will play a smaller role given that fainter and physically closer companions can be resolved. From the RK catalog alone, $25$ of the $883$ CVs that are interior to 1 kpc have wide tertiary companions ($\sim 3\%$). For those interior to $150~$pc sample, we find a CV triple fraction of $5.4\%$ ($2/37$)\footnote{A third CV triple in the RK catalog, V1108 Her, lies near the $150$~pc boundary, with median parallax-derived distances of $154$~pc for the CV and $147$~pc for its companion. Including this system would increase the CV triple fraction to $8\%$, $10\%$, and $14\%$ in the $150$~pc sample for RK, ZTF, and VSX, respectively. 
Although the companion has a smaller parallax uncertainty than the CV, and it is thus not unlikely that the source is within $150$~pc, but we exclude it to remain consistent with previous work \citep[e.g.,][]{Pala20}.}; these are V3885 Sgr and V379 Tel. For the $150$~pc sample, RK has the smallest triple fraction compared to the other catalogs, though the CVs here are almost all spectroscopically confirmed. The highest CV triple fraction in the $150$~pc sample is from the VSX catalog, potentially because it is missing some CVs in the 150 pc sample. Using the results from our $150$~pc sample, and without correcting for selection biases, we estimate a lower limit on the tertiary fraction of $10\%$ for CVs.

\subsection{Comparison to Previous Samples}\label{subsec:period_dist}
A few previous works have searched for wide companions to CVs. Notably, \citet{Pala20} generate a volume-limited sample of CVs within $150$~pc using {\it Gaia} DR2. In their sample, they identify candidate common proper motion companions to three CVs, V379 Tel, {\it Gaia }J154008.28-392917.6, and {\it Gaia }J051903.99+630340.4, but remain agnostic about whether these are gravitationally bound\footnote{V379 Tel and {\it Gaia }J154008.28-392917.6 are among the faintest X-ray sources within the 150 pc sample \citep{Rodriguez24}.}. Our sample here recovers two of them as being in bound triples, which are V379 Tel ($s=4083$~au) and {\it Gaia }J051903.99+630340.4 ($s=793$~au); the {\tt R\_chance\_align} for these systems is $5.5\times10^{-4}$ and $10^{-6}$. The separations for these are also a factor of 10 smaller than those reported by \citet{Pala20} because they estimate 3D separations which are dominated by parallax uncertainties \citep[see][their Appendix A]{Nelson2021}. Furthermore, tertiary separations of $\gsim1$~pc would likely have been unbound due to the Galactic tide. We also identify an additional CV triple in the $150$ pc sample which was not discussed by \citet{Pala20}, V3885 Sgr, with a tertiary separation of $891$~au and {\tt R\_chance\_align} = $5\times10^{-6}$. These three CV triples in the 150 pc sample are confidently bound given their extremely small {\tt R\_chance\_align}. 

Previous work has identified a common proper motion companion to the prototypical recurrent nova, T Pyx \citep[][]{Knigge22}. These authors also mention that they performed a broader search for wide companions CVs, but those results have not yet been published.
% Many previous studies suggested that observed eclipse timing variations in CV light curves could result from the gravitational influence of third bodies \citep[e.g.,][]{Martinez-Pais2000, Yang10, Chavez2012, Yang17, Chavez22}. However, recent work has shown that observed eclipse timing variations in CVs and related classes of binaries are unlikely to be driven by tertiaries, and the long-term trends may instead be more plausibly attributed to changes in the magnetic activity of the donor star \citep[e.g.,][]{Pulley22, Souza24}. 

In addition, several works have proposed the existence of tertiaries to CVs based on long-term eclipse timing variations \citep[e.g.,][]{Martinez-Pais2000, Yang10, Chavez2012, Yang17, Chavez22}. Companions inferred through this method have much shorter tertiary periods and therefore close-by companions \citep[e.g., typical periods of $10-1000$~days;][]{Yang17, Chavez22}. However, from our simulation of CV formation in triples, we find that the closest predicted tertiaries are $\sim10$~au away, making these inferred tertiaries inconsistent with any of our proposed CV formation channels (Section \ref{subsec:CV_formation} and Figure \ref{fig:CVs_a1_e1_a2}). Moreover, recent work has shown that observed eclipse timing variations in CVs and related classes of binaries are unlikely to be driven by tertiaries, and the long-term trends may instead be more plausibly attributed to changes in the magnetic activity of the donor star \citep[e.g.,][]{Pulley22, Souza24}. 

%Smadar jumped to here

\section{Triple Population Simulations}\label{sec:simulated_sample}
Using detailed three-body simulations, we aim to investigate the different formation pathways of CVs in triples. We initialize a population of $2000$ triple star systems at the zero-age main sequence (ZAMS) and evolve each triple for up to $12.5$~Gyr using our detailed three-body simulations (described below).

For CVs in triples, we also note the possibility that the inner binary evolved with little influence from the outer tertiary. 
In order to understand the role of the outer tertiary star in instigating CV formation, we separately evolve all of our inner binaries as isolated binaries using the {\tt COSMIC} binary evolution code \citep{Breivik20}. 
By comparing the isolated inner binaries to their triple counterparts, we seek to understand the importance of the outer tertiary in CV formation and explore the various dynamical histories for CVs in triples.  

\subsection{Numerical Setup}\label{subsec:numerical_setup}
\begin{figure}
    \centering
    \begin{minipage}[t]{0.49\textwidth}
        \includegraphics[width=\textwidth]{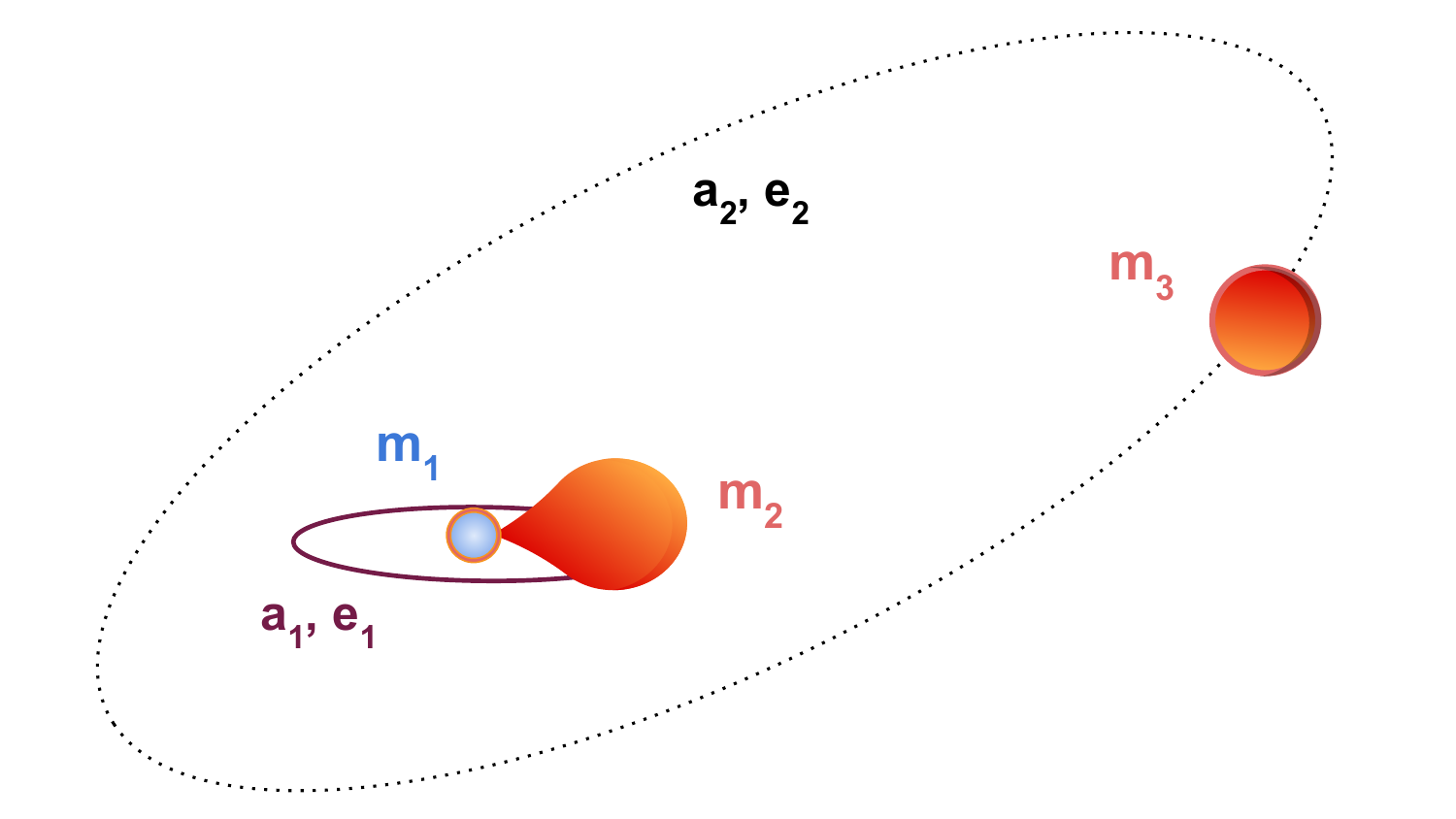}
    \end{minipage}
    \vspace{0.2cm}  % Adjust spacing between rows
    \begin{minipage}[t]{0.49\textwidth}
        \includegraphics[width=\textwidth]{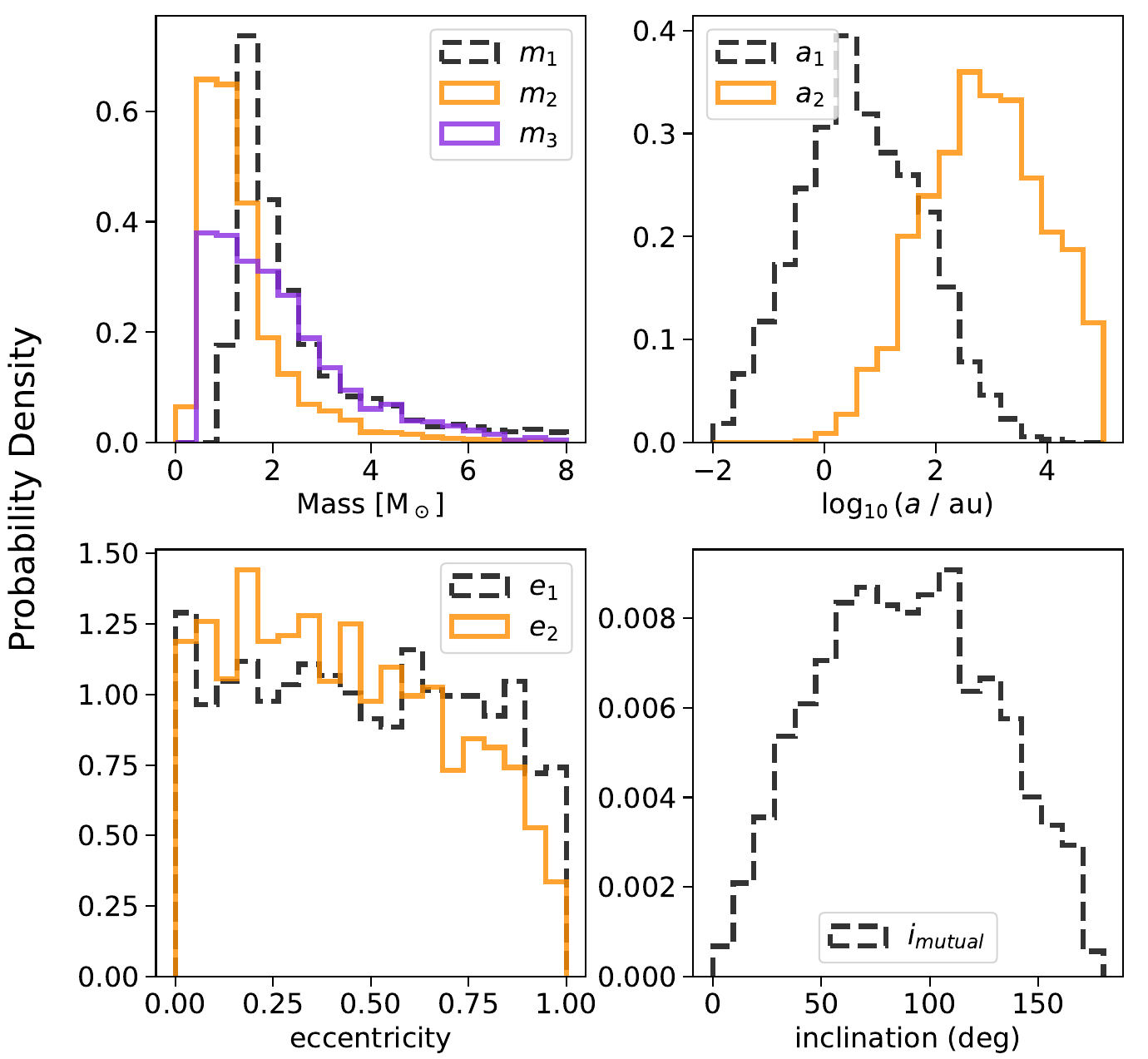}
    \end{minipage}
    \caption{{\bf Top:} Schematic representation of a triple with a CV inner binary. We label the orbital parameters, which we use throughout the paper with labeled orbital parameters. Note that this is not to scale. 
    {\bf Bottom:} Initial distributions of orbital parameters from three-body simulations. The top row plots the initial distribution for the masses (left) and semi-major axes of the inner and outer orbit (right). The bottom row shows the distribution of the initial eccentricities (left) and mutual inclination (right).
    }\label{fig:initial_distributions}
\end{figure}

We consider a hierarchical triple star system characterized by two stars with masses $m_1$ and $m_2$ forming the inner binary, while a third tertiary star (mass $m_3$) orbits the inner binary on a wider (outer) orbit. The inner (outer) orbit has a semi-major axis $a_1$ ($a_2$) and eccentricity $e_1$ ($e_2$). The triple also has a mutual inclination between the inner and outer orbit of $i_{\rm mutual}$. We label these variables in a schematic (unscaled) triple in Figure \ref{fig:initial_distributions}.

Our numerical simulations solve the three-body equations of motion up to the octupole level of approximation \citep[full set of equations in][]{Naoz2016}. We also include stellar evolution for all three stars using the {\tt Single Stellar Evolution} ({\tt SSE}) code \citep{SSE}. For both the inner and outer orbit, we include the effects of general relativistic precession to 1st post-Newtonian order \citep{Naoz2013GR}. This reasonably models the dynamics of the mass ratios considered in this study \citep[e.g.,][]{Naoz2013GR, Lim20, Kuntz22}. For the inner binary, we also track tidal effects using an equilibrium tide model, allowing for the orbit to torque the stars and vice versa  \citep[e.g.,][and see latter appendix B from the full set of equations]{Hut80,Eggleton98,Kiseleva98,Naoz2016}. Our tidal prescription changes based on the type of the star \citep{Rose+19,Stephan18,Stephan19,Stephan21}. For main-sequence stars more massive than $1.5$~M$_\odot$, we use a radiative tidal model. For low-mass main sequence stars ($<1.5$~M$_\odot$) and red giants, we assume convective tides \citep{Zahn77}. Additionally, the stars spin down due to magnetic braking, where the single-star spin-down rate is taken from  \citep{Zahn77}. 

If the inner binary crosses its Roche limit (see below), we use {\tt COSMIC} \citep{Breivik20} to model the evolution, which may include mass transfer. 
%If the inner binary begins to Roche lobe overflow, we transfer the inner binary into {\tt COSMIC} to model the mass transfer evolution. 
This effectively assumes that the inner binary is gravitationally decoupled from the tertiary during this time, which is reasonable given EKL is suppressed when tides dominate the apsidal precession of the inner binary \citep[e.g.,][]{Antognini2015, Naoz2016}. If the binary survives mass transfer and later becomes detached, we place it back into our triple code to continue its evolution. See \citet{Shariat23} and \citet{Shariat24} for details on this triple evolution process.
Simultaneously, we evolve the tertiary using single star tracks and follow the evolution of the outer orbit's semi-major axis ($a_2$) and eccentricity ($e_2$) using adiabatic prescriptions \citep[similar to][]{Shariat23}.

\subsection{Initial Conditions of Three-Body Simulations}\label{subsec:ICs}
We draw initial orbital and stellar parameters from probability distributions similar to those used in \citet{Shariat23}.  We draw $m_1$ from a Kroupa initial mass function \citep{Kroupa2001} ranging from $1-8$~M$_\odot$, where the range is chosen such that the primary will become a WD within the $12.5$~Gyr evolution time. We sample the inner binary mass ratio ($q_1$) from a uniform mass ratio distribution between $0.1$ to $1$, and set $m_2 = q_1 m_1$ \citep[e.g.,][]{Raghavan2010,Moe17}. We also sample the mass ratio of the outer binary ($q_2$) from the same uniform distribution and set $m_3 = q_2 (m_1+m_2)$. Previous triple population synthesis studies show that population-level results are relatively insensitive to the choice of mass ratio distribution \citep{Shariat23}.  The initial mutual inclinations are chosen from an isotropic distribution (uniform in $\cos i$). Observed triples with similar separations and component masses to our simulated systems show a median mutual inclination of $\sim90^\circ$ \citep[][]{Tokovinin22}, consistent with both isotropic and uniform distributions\footnote{ In contrast, compact hierarchical triples ($P_{\rm out}<1000$~days) exhibit a coplanar excess \citep[e.g.,][]{Bashi24}, though these are not the focus of this study.}.
The initial $e_1$ and $e_2$ are chosen from a uniform distribution, as are the initial spin-orbit angles for the stars in the inner binary ($\psi_1, \psi_2$). The periods of the inner and outer orbit ($P_1$ and $P_2$) are drawn from \citet{Raghavan2010} distribution, which is a Gaussian in $\log(P/day)$ with $\langle\log(P)\rangle = 5.03$ and $\sigma_{\log(P)}=2.28$. 

After sampling these initial conditions for a given triple, we require that the inner binary is not Roche crossing and that the triple satisfies dynamical stability. For the former condition, we require that $a_1(1-e_1)>R_{Roche}$ initially, where:
\begin{equation}\label{eq:roche}
    R_{Roche,ij} = \frac{R_j}{\mu_{Roche,ji}} \ ,
\end{equation}
and
\begin{equation}\label{eq:roche_mu}
    \mu_{Roche,ji}= \frac{0.49(m_j/m_i)^{2/3}}{0.6(m_j/m_i)^{2/3} + \ln(1+(m_j/m_i)^{1/3})} \ ,
\end{equation}
where $j\in{1,2}$ represents the two stars in the inner binary and $R_j$ is the radius of the star with mass $m_j$.
For the latter condition, we assess stability by applying two criteria to our initial conditions that check whether the triple is indeed hierarchical and long-term stable. To test hierarchy, we adopt the hierarchical parameter $\epsilon$, which describes the coefficient of the octupole term in the three-body Hamiltonian \citep[e.g.,][]{Naoz2013sec}
\begin{equation}\label{eq:eps_crit}
    \epsilon = \frac{a_1}{a_2}\frac{e_2}{1-e_2^2} < 0.1 \ .
\end{equation} 
To enforce long-term stability, we apply the criteria from \citet{MA2001}:
\begin{equation}\label{eq:MA_stability_crit}
    \frac{a_2}{a_1}>2.8 \left(1+\frac{m_3}{m_1+m_2}\right)^\frac{2}{5} \frac{(1+e_2)^\frac{2}{5}}{(1-e_2)^\frac{6}{5}} \left(1-\frac{0.3i}{180^\circ}\right) \ .
\end{equation}
A deviation from hierarchy does not necessarily mean that a triple becomes unstable immediately \citep{Grishin17,Mushkin20, Bhaskar21, Toonen2022, Zhang23}. However, given that all of the CV triples in our observational sample are in the very hierarchical regime, we require hierarchy in our simulations based on Equation (\ref{eq:eps_crit})\footnote{
We also evaluate the stability of our systems using the more recent stability criterion from \citet{Vynatheya22}, finding find that most ($80\%$) are stable under their algebraic framework}. In mildly hierarchical triples, quasi-secular dynamics can influence orbital evolution by breaking down the assumptions of standard, double-orbit-averaged secular theory \citep[e.g.,][]{Antonini12_quasi, Katz12,Antonini14_quasi,Luo16_quasisec,Grishin18_quasi, Tremaine23, Klein24}. However, in our simulated sample, the vast majority of systems ($>90\%$) lie  in the extremely hierarchical regime, with EKL timescales much longer than the outer orbital period (median $P_{\rm out}/t_{\rm EKL}$ = 0.005). This suggests that quasi-secular effects are unlikely to play a significant role in the evolution of most systems in our study. 
We run each triple simulation for an upper limit of $12.5$~Gyr, but systems are stopped earlier if the inner binary crosses the {\rm Roche} limit or merges entirely. 

Figure \ref{fig:initial_distributions} depicts the initial condition distributions in this study. We show the distribution of initial masses ($m_1$, $m_2$, $m_3$), semi-major axes ($a_1$, $a_2$), eccentricities ($e_1$, $e_2$), and mutual inclination ($i_{\rm mutual}$). On the top of Figure \ref{fig:initial_distributions}, we show a schematic representation of a CV in a hierarchical triple (not to scale) along with some of the key parameters labeled (masses, eccentricities, and semi-major axes). 

%Here, the Roche limit, $R_{Roche}$, is defined by 
%\begin{equation}\label{eq:roche}
%    R_{Roche,ij} = \frac{R_j}{\mu_{Roche,ji}} \ ,
%\end{equation}
%where $j\in{1,2}$ represents the two stars in the inner binary and $R_j$ is the radius of the star with mass $m_j$. $\mu_{Roche,ji}$ is the approximate Roche radius given by \citep{Eggleton83Roche}
%\begin{equation}\label{eq:roche_mu}
 %   \mu_{Roche,ji}= \frac{0.49(m_j/m_i)^{2/3}}{0.6(m_j/m_i)^{2/3} + \ln(1+(m_j/m_i)^{1/3})}
%\end{equation}

%When the inner binary begins to transfer mass, we move the inner binary from the triple code into the {\tt COSMIC} binary evolution code to track the detailed mass transfer evolution, following \citet{Shariat23, Shariat24}. Simultaneously, we evolve the tertiary using single star tracks and follow the evolution of the outer orbit's semi-major axis ($a_2$) and eccentricity ($e_2$) using adiabatic prescriptions \citep[e.g.,][]{Shariat23}.

\subsection{Binary Magnetic Braking }\label{subsubsec:MB}
For close WD+MS binaries, both before and after the onset of mass transfer, magnetic braking (MB) is one of the main angular momentum loss mechanisms that drive their evolution \citep[e.g.,][]{Verbunt81, Rappaport83}. Since CV formation and evolution are sensitive to MB, we modify {\tt COSMIC} to include the MB prescription from \citet{Rappaport83}. We note that the \citet{Rappaport83} magnetic braking prescription has several known limitations when compared to observations of both single and binary stars. Alternative models, such as saturated, disrupted, or boosted magnetic braking prescriptions \citep[e.g.,][]{Ivanova03,Matt15,EB22MB,Belloni24}, may provide better agreement with the observed CV population. However, we do not explore these alternatives here, as our focus is primarily on the role of triple evolution prior to the common envelope phase. A comprehensive evaluation of different magnetic braking models and their impact on CV formation is beyond the scope of this work but is well warranted in future studies.

As we discuss with detail in Appendix \ref{app:MB}, the updated MB model leads to consistent evolution between the {\tt COSMIC} binaries and {\tt MESA} binary evolution for CVs, up to the period gap. Before changing the MB, many of the inner binaries in triples would merge almost instantly after mass transfer. Now, with the new MB prescription, binaries remain stably mass transferring for longer periods which matches {\tt MESA} (Figure \ref{fig:MESA_vs_COSMIC_MB}), as is consistent with observed populations of CVs \citep[e.g.,][]{Rappaport83}. Note {\tt COSMIC} merges the binary earlier than {\tt MESA} predicts. 
%Note, however, that {\tt COSMIC} still merges the binaries at seemingly arbitrary times in many of the binary models. 
Nevertheless, since our study focuses mainly on the properties at CV formation (i.e., first mass transfer), the subsequent CV evolution via {\tt COSMIC} is not essential to our conclusions.  
%of the CV is not very important for our results. 

\section{Results}\label{sec:results}
\subsection{Outcomes of Population Synthesis}
% Insert Population Statistics (maybe add a table).
To identify the CVs among our population of simulated triples, we select inner WD+MS binaries that are stably mass transferring. Here, `MS' stars exclude stripped helium stars on the MS and Hertzsprung gap stars, and we only consider those with {\tt SSE/BSE} star type {\tt kstar} = 0 or 1 (corresponding to low-mass stars with $M < 0.7$~M$_\odot$ and $M = 0.7 - 8$~M$_\odot$). Overall, we find that $47/2000$ of the triples harbor inner binaries that are CVs according to this criteria. While this fraction may seem high, we remind the reader that our triple simulations did not include primary stars with $m_1<1$M$_\odot$. The implications from our model on the number of triples in the galaxy are discussed in Section \ref{subsec:N_CVs}.
We focus on characterizing their triple orbits at the first moment where the inner binary became a CV. The median time of CV formation in triples is $1.7$~Gyr with a standard deviation of $3.8$~Gyr. The CVs have periods between $3$~hrs and $2$~days. Note again that these are the periods at the {\it onset} of WD+MS mass transfer, so they are expected to decrease on average through mass transfer, magnetic braking, and gravitational waves. Also, since CVs evolve more rapidly at long periods than at short periods, observations of the simulated systems at random times would produce a period distribution with more emphasis at short periods.

If we expand our search to include all types of non-compact object companions that are stably mass-transferring systems with a WD, such as giant and helium star donors, then we find $170/2000$ systems. This means that there are $\sim3$ times more mass transfer interactions between WD+evolved companion inner binaries than there are for WD+MS inner binaries\footnote{These triple-induced WD mass transfer rates are similar to those predicted for binaries in the Galactic Center \citep[e.g.,][]{Stephan19}.}. While these binaries would not be classified as CVs, they are expected to evolve into related types of systems, primarily symbiotic (red giant + WD) and AM CVn (helium star + WD) binaries. This large fraction of evolved donors transferring mass with a WD is not surprising, given that evolved companions have larger stellar radii, which increases the chance of interaction and the efficiency of tides to shrink the inner orbit. 
As we discuss further in Section \ref{subsec:CV_formation}, we expect the triple channel to preferentially produce CVs with more evolved donors and longer orbital periods than the standard binary channel for some fraction of the CVs. In Appendix \ref{app:WD_accreting} and Figure \ref{fig:CVs_a1_e1_a2_MT2}, we outline the characteristics of accreting WD systems with evolved donors, though the evolution of these systems in triples warrants further detailed study.

\begin{figure*}
    \centering
    \includegraphics[width=1.0
    \textwidth]{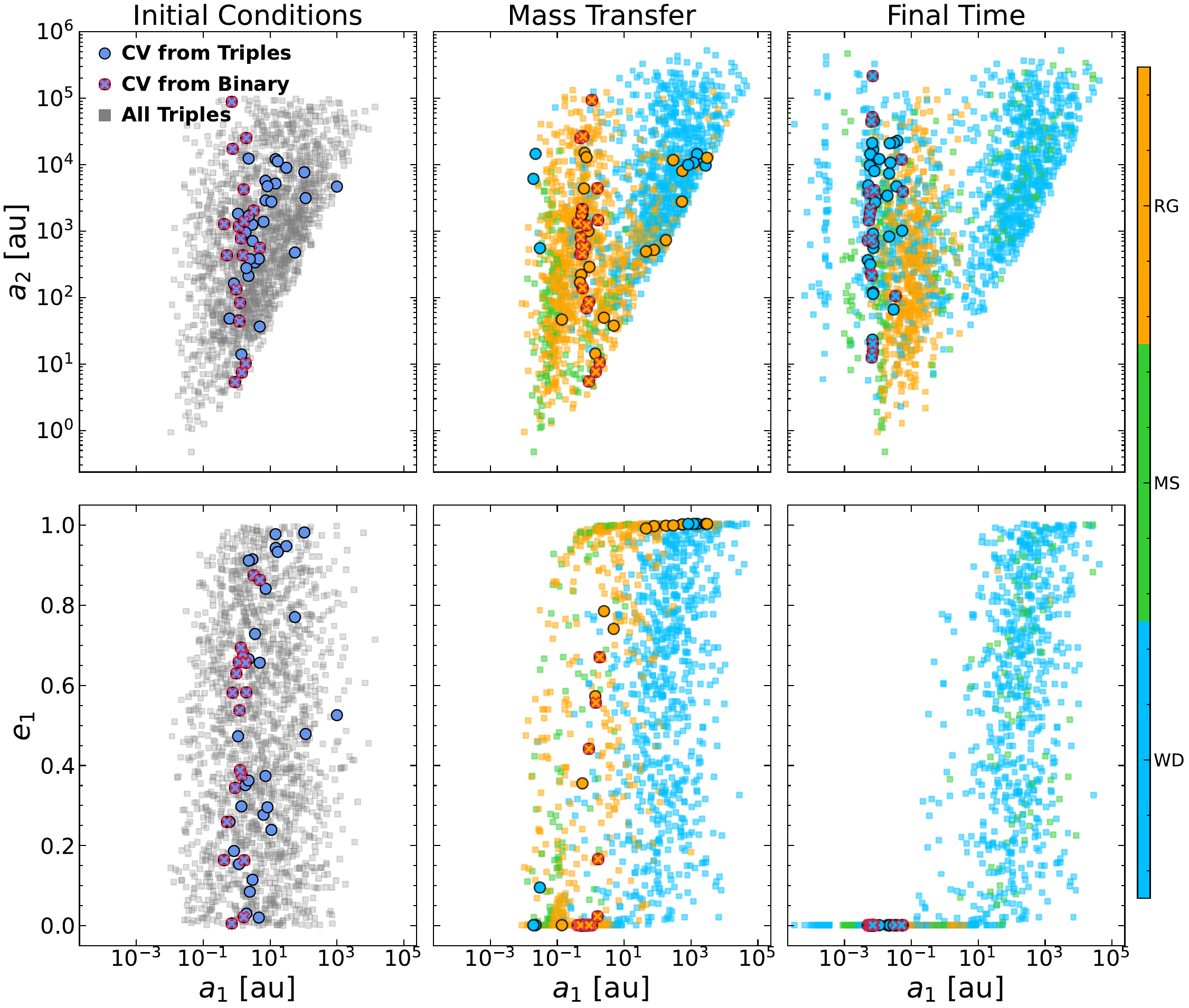}
    \caption{Orbital parameters of CV triples at different stages of evolution. Only triples that eventually become CVs are marked with larger points circular points.
    In the first column, we show the initial conditions of triples in our population. The blue outlined points eventually become CVs, and those with red crosses inside of them become CVs even in isolated binary evolution simulations. In the second and third panels, we show the orbital parameters of all triples at first mass transfer (including those that do not become CVs) and at the final simulation time, respectively. In these panels, we color the points by the stellar type of the primary, which can either be main-sequence (MS), red giant (RG), or white dwarf (WD). The final time is defined as (1) the time of CV formation if the inner binary became a CV, or (2) the time right before the merger if the inner binary merged, (3) $t=12.5$~Gyr otherwise. To understand the triple evolution of all WD accreting inner binaries, see Figure \ref{fig:CVs_a1_e1_a2_MT2} in Appendix \ref{app:WD_accreting}.
    }\label{fig:CVs_a1_e1_a2} 
\end{figure*}

\subsection{Formation of CVs in Triples}\label{subsec:CV_formation}

In the canonical isolated binary CV formation model, CVs form after at least one common envelope (CE) event \citep[e.g., see][]{Zorotovic20}. The primary star in the MS+MS progenitor binary evolves into a red giant (RG) and engulfs the companion, leading to a CE phase. If the donor successfully ejects the envelope, the stars emerge as a tight but detached WD+MS binary. At this stage, the orbital angular momentum and energy are dissipated by MB and gravitational waves, shrinking the binary and creating a CV. For CVs in triples, this formation pathway can still take place when the tertiary has minimal dynamical influence on the inner binary. However, we find that in most cases, the presence of a distant third star alters the evolution of the CV progenitor, and this evolution deviates from the classical picture of CV formation. CVs in triples form through several new pathways that are otherwise impossible in isolated binaries. The main CV formation channels can be summarized as:
    \begin{enumerate}
        \item {\it High Eccentricity Formation (no CE):} 
        The CV begins as a wide binary ($\gsim10$~au) where the white dwarf forms without ever transferring mass to the secondary. Then, EKL oscillations excite the binary's eccentricity, allowing tides and mass transfer to shrink the orbit, which initiates Roche lobe overflow, forming the CV. 
        \item {\it High Eccentricity Formation (with CE)}: Prior to WD formation, wide CV progenitors are driven into sufficiently eccentric orbits, causing mass transfer within the inner binary, most often when the primary is a RG. This mass transfer can begin after the binary circularizes and the primary evolves to fill its Roche lobe, or while the orbit is highly eccentric.
        \item {\it Classical CV Channel}:
        The gravitational influence of the tertiary is small, and the inner binary becomes a CV through the classical binary channel, including a CE phase. 
        \end{enumerate}
We investigate these CV formation pathways and their relative occurrence rates by comparing the results of our triple population synthesis to both our observations and our isolated binary population.

In Figure \ref{fig:CVs_a1_e1_a2}, we display the orbital configuration, specifically, $a_2$ and a function of $a_1$ (top row) and $e_1$ vs. $a_1$ (bottom row) of triples at different stages of their evolution. In the left column, we consider the initial configuration when all three stars are on the ZAMS. The larger points show triples where the inner binary eventually becomes a CV. The CV points with red crosses indicate those binaries that also became CVs in our isolated binary evolution models (these represent $38\%$ of all CV systems). These systems would have become CVs even without the presence of the tertiary, although their detailed evolutionary histories are often different between the triple and isolated binary simulations. Only $20\%$ of these systems formed without high eccentricities, through the classical CV channel. The inner binaries that also became CVs without a tertiary present span a narrow range of initial binary separations around $1$~au. On the other hand, the systems that only became CVs in triples ($62\%$ of all CVs) extend to wider initial binary separations, $a_1$. For these systems, gravitational effects of the tertiary through the EKL mechanism were essential to their eventual evolution into CVs. We describe these outcomes below.
% Over secular timescales, the inner and outer orbits exchange angular momentum, leading to eccentricity and inclination oscillations to the inner binary through the EKL mechanism \citep{Naoz2016}. Combined with stellar evolution and tides, high eccentricity excursions ($e_1>0.99$) during EKL can allow wide inner binaries (up to $1000s$ of au) to shrink and interact through tides and mass transfer, which will be significant given their short periastron distances during this high eccentricity stage. A similar formation mechanism is crucial for many black hole low-mass X-ray binary population \citep[][]{NaozLMXB,Shariat24c}. 

The middle column shows the point of evolution where mass transfer first took place. Therefore, it shows systems at different evolutionary times. The color code here represents the stellar type of the more massive progenitor star (see Section \ref{subsubsec:without_CE} and Section \ref{subsubsec:with_CE} for further discussion about this stage). %For example, this panel highlights that many (X) WDs do not undergo any mass transfer until this stage. ...
As highlighted in the bottom middle panel, while most CV triple formation pathways entail high ($e_1>0.9$) eccentricities during the binary evolution, the specific formation pathways can vary between different CVs. For example, this high eccentricity event can initiate tides and mass transfer when the primary is an MS, RG, or even after it has already become a WD. We, therefore, focus on the onset of mass transfer in the inner binary to uncover the different formation pathways for CVs in triples.

The last column of Figure \ref{fig:CVs_a1_e1_a2} shows the triple configuration at the end of their evolution. We define the `end' of a triples evolution as $t=12.5$~Gyr for the detached triples. If the inner binary became a CV, we instead plot the orbital parameters at the time of CV formation. If the inner binary merged and never was a CV, we define the final time as the time right before the merger event took place. Most orange and green points in this final panel are mergers, while the primary was an MS or RG. All of these have circular orbits just before the merger due to {\tt COSMIC}'s instantaneous circularization upon mass transfer, although many originated from a highly eccentric inner orbit. Such merged systems are studied in detail in \citet{Shariat24}. This panel also contains systems where a WD is accreting from a companion that is not a MS star; we highlight these systems in Figure \ref{fig:CVs_a1_e1_a2_MT2} and Appendix \ref{app:WD_accreting}.

\subsubsection{Channels Without a Common Envelope Phase}\label{subsubsec:without_CE}

As depicted in the second column of Figure \ref{fig:CVs_a1_e1_a2}, there are many ($7$) CVs that never experienced a common envelope phase (large blue points). Namely, their first mass transfer interaction occurred when the primary was already a WD. This constitutes $15\%$.
At the time of mass transfer, three of these had already been circularized partially or entirely due to tides. The other four begin their WD+MS mass transfer on wide ($a_1\sim10^3$~au) eccentric ($e_1\sim0.9999)$ orbits. At this stage, dissipative forces, such as tides and mass transfer, remove orbital energy and angular momentum from the binary, causing them to shrink and circularize, becoming CVs (rightmost panel of Figure \ref{fig:CVs_a1_e1_a2}). These CVs form without ever experiencing a common envelope phase.

While the number of these CV outcomes here is relatively small, the fraction is consistent with rates of similar mass transfer channels reported in previous triple population studies \citep[e.g.,][]{Toonen2016, Stephan19, Shariat23}. Thus, simply applying Poisson error bars to these fractions would likely overestimate the uncertainty, and these results are robust. We chose not to simulate a larger population because we focused on modeling the full evolution for all 2000 triples to get an accurate and complete population, which required over 50,000 CPU hours. Approximately $30\%$ of these triples required more than 24 hours to converge, and some simulations required up to a week to complete\footnote{We note that triple integration techniques often impose a shorter integration time limit. Such a practice misses the majority of these systems.}.

\subsubsection{Channels With a Common Envelope Phase}\label{subsubsec:with_CE}
Most of our simulated triples ($40/47$) do experience a common envelope event prior to becoming CVs. However, the common envelope channel is enhanced thanks to the tertiary eccentricity excitation.
%Among them, 
Specifically, $18\%$ ($7/40$) begin mass transfer on wide, eccentric orbits ($a_1\gsim10$~au and $e_1>0.9$). This population comes from the initially wide inner binaries that had their eccentricities excited by the tertiary to the point where their pericenter partially fills the companion's Roche lobe. Five are RG+MS, and two are MS+MS with eccentricities near $e_1\sim0.99$. At this stage, dissipative forces, such as tides, rapidly shrink the orbit, initiating dynamically unstable mass transfer and a subsequent common envelope phase. Post CE, the WD+MS pre-CV shrinks due to magnetic braking, eventually becoming a CV.

A very similar population of CV-forming inner binaries are those that also started wide but had their orbits shrink and circularize before the onset of mass transfer. These systems also experience a common envelope phase, except for the three systems that already had WD primaries. Circularization prior to mass transfer (or CE) occurs in ($24/40$) of the CE CV-forming population and is shown in the bottom of Figure \ref{fig:CVs_a1_e1_a2}'s `Mass Transfer' column with $e_1=0$. Most of these ($19$) are binaries that tidally lock when the primary first becomes a red giant (RG+MS binaries). These are similar to the wide ($a_1>10$) RG+MS binaries that are radially mass transferring; the only difference is that these are circularized first, and then the primary evolved to fill its Roche Lobe. Note that these WD+MS also avoided a common envelope with the WD progenitor. 
% Find the number of systems that experience a common enevlope in the first place and write it here and maybe in a table
% make a table that splits when the first mass transfer was and maybe even by the eccentricity
Additionally, many of the RG+MS CV progenitors have moderate eccentricities at the start of mass transfer ($e_1 = 0.2-0.8$). The systems all have pericenter distances of $\sim1$~au, which is inside the Roche lobe of the evolved primary. In fact, at the onset of mass transfer, most inner binaries contain at least one red giant ($73\%$). The larger radius of an evolved giant makes tidal capture or mass transfer more likely in the inner binary, especially in triples where the eccentricity can become excited during EKL cycles.

\subsubsection{Comparison to Isolated Binary Models}\label{subsubsec:with_CE}
Using {\tt COSMIC}, we initialize a binary population where each system is one of the inner binaries from our triple population (2000 total binaries). Modeling these binaries without a tertiary will allow us to constrain the role of a tertiary in CV formation. We find that $3.5\%$ of these isolated binaries formed CVs, compared to $2.5\%$ of inner binaries that become CVs in the triple channel. We address these efficiencies below. 

All of the CVs formed in the isolated binaries channel had initial separations of $\sim1-5$~au and formed through the classical CV channel, which entails a common envelope phase.
At face value, assuming identical initial conditions, removing the tertiary star increases the number of CVs. This is because, in triples, strong EKL interactions can excite high eccentricities and cause some systems to merge at an earlier stage of their triple dynamical evolution \citep[e.g.,][]{Weldon24}. In fact, the systems that merged in triples but became CVs in the isolated binary population had a median initial tertiary separation of $111$~au and an initial EKL timescale of $7$~Myr. In contrast, the CVs that formed in triples have a median initial tertiary separation of $1165$~au and an initial EKL timescale of $800$~Myr.

However, this exercise misses the wide isolated binaries because such wide binaries would rarely fit in a stable hierarchical triple and are thus missing from this isolated binary population. To account for the wide isolated binary population, we initialize another {\tt COSMIC} binary population with the same masses and eccentricities as the isolated binaries, but now choose the initial orbital periods from a general \citet{Raghavan2010} binary distribution (not restricted to any triple stability hierarchy criteria). 

From this population, we find that $1.25\%$ ($25/2000$) of binaries become CVs. This is lower than the $3.5\%$ formed from the previous isolated binary population, which were the inner binaries of triples. Therefore, the triple channel is two times more efficient at producing CVs at the masses and separations considered here ($2.5\%$ compared to $1.25\%$). In practice, the efficiency of either channel depends also on the binary vs triple fraction of stars in the Galaxy. Assuming that there are three times more binaries as triples \citep[e.g.,][]{Moe17}, then the triple channel would account for $40\%$ ($2.5/(2.5+3.75)$) of all CVs the Galaxy.

\subsection{Number of CV triples in the Galaxy}\label{subsec:N_CVs}
Using our simulation results, we estimate the number of Galactic CVs in triples. We apply the following equation,
\begin{equation}\label{eq:N_CV}
    N_{\text{CV, triple}} = 
    \tau_{\text{CV}} \times 
    \text{SFR}\times
    f_{m_1>1}\times  
    f_{\text{triple}}\times 
    f_{\text{CV}} \ ,
\end{equation}
where $N_{\text{CV, triple}}$ is the number of CV triples in the Galaxy, $\tau_{\text{CV}}$ is the period of time over which the binary is observable as a CV, $\text{SFR}$ is the star formation rate (in units of stars per year), $f_{m_1>1}$ is the fraction of stars with a mass greater that $1$~M$_\odot$ (since we assume this for $m_1$ in our simulations), f$_{\text{triple}}$ is the fraction of stars that reside in triples, and $f_{\text{CV}}$ is the fraction of triples that host CV inner binaries in our simulations. For this estimate, we set $\tau_{\text{CV}} = 1$~Gyr, $\text{SFR} = 2$~yr$^{-1}$ (roughly corresponding to a Milky Way SFR of $1$~M$_\odot$~yr$^{-1}$), $f_{m_1>1} = 0.1$, f$_{\text{triple}} = 0.1$, $f_{\text{CV}} = 47/2000 = 0.0235$. Under these rough order-of-magnitude assumptions, we estimate $4.7\times10^5$ CV triples to reside in the Galaxy today. Note that this estimate is sensitive to the initial conditions assumed in our triple simulations, and different initial separations and mass ratios will change this calculation. The value of $N_{\text{CV, triple}}$ is a factor of $5-10$ smaller than the total number of CVs estimated to reside in the Milky Way \citep[e.g.,][]{Pala20, Rodriguez24}, which is consistent with our observationally-inferred lower limits. 

\begin{figure*}
    \centering
    \includegraphics[width=0.99
    \textwidth]{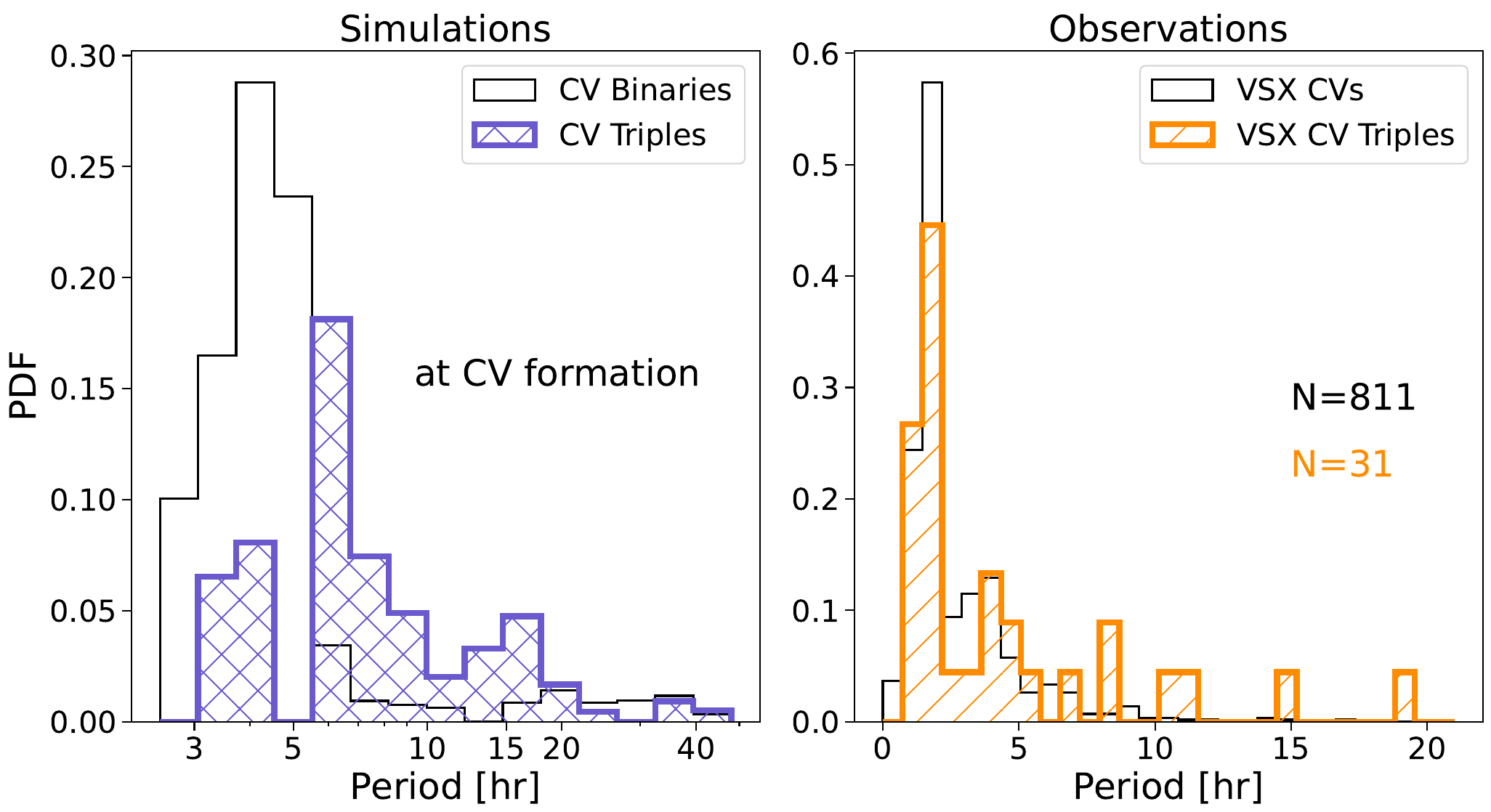}
    \caption{Period distribution of CVs with (purple) and without (black) tertiary companions. The histograms are normalized such that the area under the curves is 1. {\bf Left:} We compare the period distribution from our isolated binary simulations and our triple simulations, where the x-axis is in log scale. We show the periods at the time of CV formation. Many of the CVs in triples originate from wide binaries that previously had high eccentricities but have since shrunk due to tides and mass transfer. This route prefers larger MS companions, which leads to mass transfer at longer orbital periods. {\bf Right:} Observed VSX CVs with (orange) and without (black) resolved {\it Gaia} wide companions. Note that these are, in most cases, observed long after CV formation and, thereby, have shorter orbital periods than the simulations, which show the period at the start of mass transfer. There are $811$ VSX CVs with reported periods, and $31$ of them have wide tertiary companions. The CV triples show a larger fraction of systems with wider orbital periods. This is consistent with these systems forming from the triple channel, with wide progenitor binaries (see Section \ref{subsec:CV_formation}).}\label{fig:CV_periods}  
\end{figure*}

\subsection{CV Period Distribution}\label{subsec:CV_periods_sims}
At this final stage, the mass of the CV has first formed, and all of the CVs are circularized with short ($a_1<0.1$~au) orbits and a range of outer tertiary separations ($a_2$). Interestingly, the $a_1$ distribution of CVs is roughly bimodal (Figure \ref{fig:CV_periods}), with $26\%$ of systems having inner separations around $5$~R$_\odot$ ($14-24$~hr periods) and the other $74\%$ having more-typical CV separations of $1.2$~R$_\odot$ ($3-10$~hr periods). The wider CV population, representing a fourth of our sample, corresponds to more evolved or massive MS donor stars, which have larger radii and fill their Roche lobe for longer periods. 
% These wider CVs have a median donor mass of $2.3$~M$_\odot$, while the shorter population has a typical donor mass of $0.7$~M$_\odot$.

In our sample of CV triples, three systems clearly resemble these wide CVs: IGR J19308+0530822, SWIFT J2124.6+0500853, and ATO J145.8742+09.1629. ATO J145.8742+09.1629 is a confirmed CV with a period of $19.46$ hours and a tertiary that is $1178$~au away, making its architecture similar to several of the simulated CVs. The other two similarly wide accreting systems, IGR J19308+0530822 and SWIFT J2124.6+0500853 were previously classified as X-ray binaries in VSX. However, a closer look shows that they likely host WD accretors rather than NSs and are, therefore, CVs (Appendix \ref{app:IGR}). They have orbital periods of $\sim15$ and $\sim20$ hours, respectively. While magnetic braking would still shrink their orbits, these stars evolved more rapidly than a prototypical CV and would likely still experience stable thermal-timescale mass. Depending on the evolutionary state of the donor, some will eventually evolve toward long periods and form detached low-mass WD binaries. In contrast, others will become evolved CVs and AM CVn binaries \citep[e.g.,][]{Podsiadlowski03, EB21CV}.

\begin{figure*}
    \centering
    \includegraphics[width=1.0
    \textwidth]{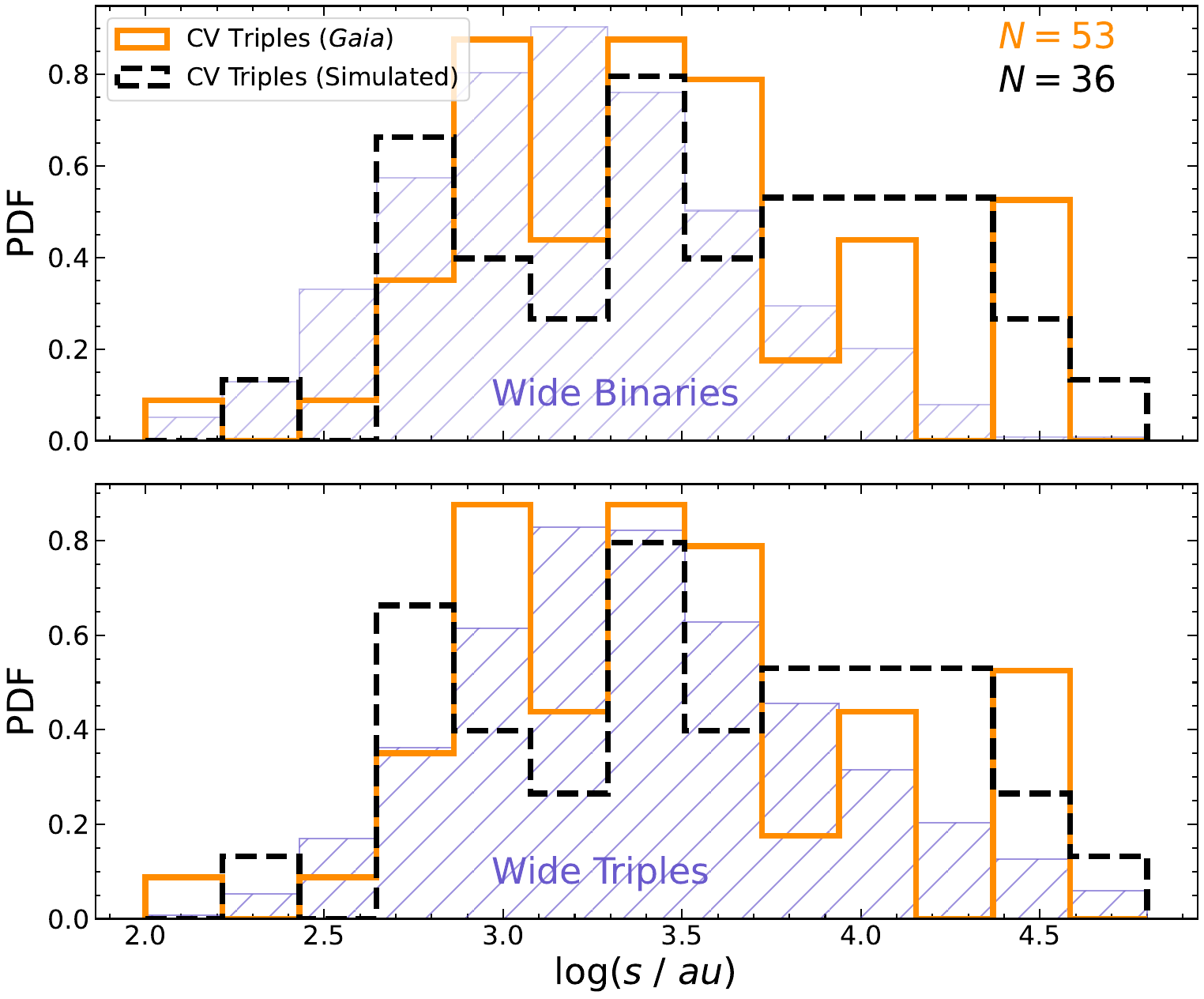}
    \caption{Separations distribution of observed CV triples (orange), simulated CV triples (black) compared to wide binaries (purple, top) and wide triples (purple, bottom) in the field. {\bf Top:} The orange distribution shows the separations of {\it Gaia} CV triples (between the CV and the tertiary star), and the purple shows the distribution for {\it Gaia} wide binaries from \citet{EB21}. In black, we plot the separations of CV triples from our triple population simulations. The simulated CV triples in this figure are only those with tertiaries that would be resolvable by {\it Gaia}. 
    {\bf Bottom:} Identical to the top except the background distribution is wide {\it Gaia} triples from \citet{Nagarajan24}, where the inner binary has an astrometric {\it Gaia} solution. Since these triples have inner binaries with $a_1\sim1$~au, some of them are progenitors of CV triples and, therefore, have more consistent distributions with our simulated sample.
    }\label{fig:sep_hists_dist} 
\end{figure*}

\subsection{Tertiary Separation Distribution}\label{subsec:sep_distribution}
We plot the tertiary separation distribution of CV triples in Figure \ref{fig:sep_hists_dist}. For the observed sample of CV triples (orange), we plot the projected separation between the CV and the wide tertiary companion. For the theoretical population of CV triples (black), we plot the semi-major axis of the outer orbit ($a_2$). In the purple histogram, we plot the separation distribution of {\it Gaia} wide binaries (top row) or the {\it Gaia} wide triples (bottom row), most of which have main sequence components. The wide triples catalog from \citet{Nagarajan24} is a subset of the wide binaries from \citet[][]{EB21} for which one component has a non-single star astrometric solution from {\it Gaia} DR3, with a typical inner separation of 1 au. Therefore, some of these systems are likely progenitors of CV triples, so we expect these to have a similar outer separation distribution to CV triples, which they do (Figure \ref{fig:sep_hists_dist}, bottom row). We focus on comparing the observed CV triples to the observed wide binaries (first panel of Figure \ref{fig:sep_hists_dist}), which will highlight differences between these two populations. 

To compare the observations to simulations, we plot the simulated CV triples that would be resolved by {\it Gaia}, which we determine based on a mock observation routine. To simulate {\it Gaia} observations, we place each of our simulated CV triples at a randomly selected distance that we sample from the distances of the observed CV triples. Based on this distance, we check whether the tertiary would be resolved at its given physical separation, which we take to be the outer orbit's semi-major axis ($a_2$), assuming an angular resolution of 1.'' We repeat this procedure five times for various distances and keep those CV triples that would be observed for more than three of the sampled distances. Out of the $47$ CV triples, $35$ of them would be resolved with {\it Gaia} at the distances of our observed sample. We note that the semi-major axis of the tertiary orbit is different from its projected separation at the time of observation, which might affect the simulated distribution in Figure \ref{fig:CVs_a1_e1_a2}. However, this discrepancy is usually of order unity \citep[e.g.,][]{Dupuy11,EB18}.

We discover that {\it Gaia} CV triples are wider than regular {\it Gaia} wide binaries (Figure \ref{fig:sep_hists_dist}, top) and slightly wider than the {\it Gaia} wide triples (Figure \ref{fig:sep_hists_dist}, bottom). Comparing this to separations from the simulated CV triples (black curve), we similarly find that they are wider on average. Both the observed and simulated CV triples have an excess of tertiary separations around $10^3-10^5$~au. The preference for wider orbits is an imprint of the formation of CVs in triples. Specifically, we show from our simulations that a substantial fraction of the CVs that form in triples began as wide MS+MS inner binaries ($10-1000$~au). For triple stability to hold for these wide initial inner separations, the tertiary must be correspondingly wider ($10^3-10^5$~au), which can explain the wide tertiary separation in both our simulated and observed populations. These CVs that began as wide binaries would not have become CVs in isolation. Additionally, even the CVs that would have become CVs as isolated binaries have $a_1\sim 1$~au, which requires a reasonably wide outer orbit and removes some of the smaller $a_2$ triples from the sample, resulting in a wider outer separation on average.

In summary, the separation distributions (Figure \ref{fig:sep_hists_dist}) reveal three main findings. First, CV triples tend to be wider than typical wide binaries because the outer orbit must accommodate an inner binary of at least 1 au. Second, the tertiary separations of CV triples are similar to those of wide astrometric triples, some of which may eventually evolve into CVs. Finally, population-level simulations largely reproduce the observed wide nature of CV triples.

\section{Discussion}\label{sec:discussion}
We aim to construct the first statistical sample of CVs in triples and study their formation channels, which often deviate from traditional CV formation histories. However, both our sample and previous efforts searching for common proper motion companions {\it Gaia} are only sensitive to the widest, resolved tertiaries. From simulations, we predict that $30\%$ of the tertiaries to CVs triples are within $500$~au (Figure \ref{fig:CVs_a1_e1_a2}), making them difficult to detect with {\it Gaia} at $d > 500$~pc. Other techniques that can resolve or infer closer companions need to be applied to account for this missing sample, such as high-resolution imaging. Previous imaging results find a deficit of close ($2-100$~au) tertiary companions to nearby CVs \citep[][]{Shara21}. In future work, we aim to further explore the presence of tertiary companions at closer projected separations using high-resolution imaging. 

 In this study, we do not self-consistently track eccentric mass transfer, which is especially important in three-body evolutionary scenarios. In our {\tt COSMIC} binary simulations, the inner binary is circularized as soon as mass transfer begins, leading to qualitatively different outcomes compared to other eccentric mass transfer models \citep[e.g.,][]{Sepinsky09,Hamers19_eccentricMT, Glanz21_eccentricCE, Rocha24}. \citet{Knigge22}, who identified a wide proper motion companion to the recurrent nova T Pyx, showed that highly eccentric mass transfer might play a role in such short-period recurrent novae. Note also that some of our outlined formation channels (Section \ref{subsec:CV_formation}) are similar to the proposed channels for T Pyx. Future models with self-consistent eccentric mass transfer would be more accurate in following CV evolution, especially in triples where high eccentricities are common.  Future models could also explore different angular momentum loss and mass transfer stability prescriptions, which may change the fraction of systems in a population that evolve into CVs \citep[e.g.,][]{Belloni18_popsynth}. One relevant change for modeling CVs is incorporating consequential angular momentum loss (CAML) \citep{King95,Schreiber16_CAML}. In this work, however, our aim is to broadly model CV formation pathways in triple systems up to the onset of the first mass transfer phase.   

White dwarf natal kicks from asymmetric mass loss on the asymptotic giant branch (AGB) would naturally widen the tertiary separations in the CV population. We do not consider the effects of mass loss kicks on the inner and outer orbits in our triple in our study. Imprints of WD kicks have been proposed in wide double WD binaries \citep[][]{EB18} and WD triples \citep{Shariat23}. Additionally, for the CVs that form through CE,  the mass loss experienced during the spiral-in phase is generally rapid compared to the orbital timescales. In this case, the mass loss could provide an impulsive kick to the system, potentially widening or unbinding the tertiary star \citep[e.g.,][]{Alcock86, Toonen17,Michaely19CE,Igoshev20}. Early works \citep{Michaely19CE} proposed using wide companions to constrain CE timescales and mass loss, which was later pursued statistically by \citet{Igoshev20}. While the physics and timescales of CE evolution remain uncertain \citep[e.g.,][]{Ivanova13}, previous studies report that the timescale for the CE phase is generally below $10^5$~yrs for WD+MS binaries \citep[e.g.,][]{Ivanova13,Michaely19CE, Igoshev20}. Therefore, rapid or asymmetric mass loss can potentially generate a small impulsive kick within the triple during WD formation and/or the CE phase, if one occurs. This kick, although modest, is often comparable to the orbital velocity of CV tertiary companions since most tertiaries are wide. Such a kick would unbind a fraction of the CV tertiaries, rendering them no longer bound to the CV and thereby no longer observable as CV triple. Note that since the tertiary was bound up until WD formation (or the first CE phase), secular three-body dynamics could still be influential to the system's pre-mass transfer evolution.
With or without WD kicks, we expect CVs with wide WD tertiaries to exist, with most faint WD companions being difficult to detect with {\it Gaia}. 

In Figure \ref{fig:sep_hists_dist}, we conservatively assume the effects of adiabatic mass loss, which widens the triple orbits in proportion to the mass lost from the system \citep[see also ][for early considerations of mass-loss into secular evolution]{Kratter12,Shappee13,Michaely2014}. There also exists a mass loss regime between the adiabatic and impulsive approximations, in which both the semi-major axis and eccentricity can change. We do not expect this scenario to change our results significantly since considering both limiting cases (i.e., adiabatic and impulsive) did not change our main conclusions (Appendix \ref{app:impulsive_ML}). In CV triples with WD tertiaries, there are two total WDs in the triple system suggesting two episode of mass loss in the triple. Therefore, the tertiary was likely closer to the CV in the past considering that the orbits widened during mass loss, implying a stronger gravitational influence on the CV progenitor binary during the history of the triples evolution.

In Appendix \ref{app:impulsive_ML}, we consider the impact of impulsive mass loss on resulting separations of CV-tertiaries. The physical mechanism for this is kept agnostic, but may apply to WD natal kicks or CE kicks due to rapid and/or asymmetric mass loss. In summary, we find that such kicks shift the entire distribution by widening most binaries and ionizing some (Figure \ref{fig:sep_kicks}). Assuming mass-loss kicks would suggest that an even larger fraction of isolated CVs formed through triple dynamics pathways, and their tertiary became unbound during the CV mass transfer evolution or white dwarf formation.  Interestingly, the presence of a significant number of CVs with bound tertiary companions at separations exceeding $10^3$~au supports the idea that common envelope (CE) evolution does not always impart strong kicks, since some fraction of these systems likely experienced a CE phase. At such large separations, the orbital velocity of the tertiary is $\sim1$~km s$^{-1}$, suggesting that any kick imparted due to rapid mass loss during the CE phase must be relatively small, likely no more than a few km s$^{-1}$. This result supports models where CE `kicks' are not large or where the CE phase is not too rapid (relative to orbital timescales), such that wide triples can remain bound post-CE \citep[e.g.,][]{Sandquist98, Shiber19, Chamandy19, Michaely19CE, Igoshev20}. It may also be the case that all of these formed without a CE, though our simulations support that a fraction of them did. Another scenario where CVs can form without a CE phase is in dense cluster environments, where strong dynamical encounters and exchange can occur \citep[][]{Knigge12_GCs,Belloni16_GCs, Belloni17_GCs,Belloni21_cluster,Kremer20}. However, recent observations suggest that dynamical interactions in globular clusters cannot efficiently form CVs \citep[][]{Rivera24_GCs}.

We note that the separation distribution of the simulations will be susceptible to the initial separation distribution (Figure \ref{fig:initial_distributions}). At the same time, the mass transfer outcomes are not highly sensitive to initial separations distributions \citep[see, for example,][]{Shariat23}. Even if we only compare the CV triples to their wide binary counterparts, both of which are susceptible to similar systematics, we still observe a robust excess of wide CV triples.

The triple formation pathways proposed for CVs are remarkably similar to those proposed for the formation of black hole low-mass X-ray binaries (BH-LMXBs) \citep[][]{NaozLMXB, Shariat24c}. Both classes of binaries involve a main-sequence (MS) donor star transferring mass onto a compact object— a white dwarf (WD) in the case of CVs and a black hole (BH) in the case of BH-LMXBs. In both cases, tertiary companions excite extreme eccentricities ($e_1\sim0.9-0.9999$) to wide compact object+MS inner binaries, which can initiate mass transfer events before or after tidal circularization. Similar to the CVs, the BH-LMXB progenitors may be placed on these near-radial orbits either before the BH forms, in which case a CE phase may ensue, or after, in which case the mass transfer between the BH+donor star produces a BH-LMXB \citep[][]{NaozLMXB, Shariat24c}. In contrast to a CV common envelope, the larger mass ratios between a BH progenitor and a low-mass MS companion might make successful envelope ejection more difficult in a BH-LMXB progenitor binary. Observationally, testing the common envelope scenario may be possible by using molecular abundance ratios as tracers \citep[e.g.,][]{Sarna95, Dhillon02}. 
Interestingly, a BH-LMXB in a hierarchical triple was recently observed \citep{Burdge24}, confirming that BHs can form in triple environments. Overall, in both CVs and BH-LMXBs, triples provide a unique formation route in which the such can form without a CE phase. CVs are also more numerous than BH-LMXBs, allowing us to study their population demographics in more detail and distinguish between their various formation models. Insights from the CV population can potentially be used to learn more about BH-LMXB formation and evolution.

\section{Conclusion}\label{sec:conclusions}
This study investigates the formation and evolution of CVs in triple-star systems. Our main conclusions are summarized as follows.
\begin{enumerate}
    \item {\bf  CV Triples Observed Catalog}: We construct a sample of $\sim50$ CVs in triples. Our sample includes wide, resolved tertiary companions to CVs within $1$~kpc of the Sun using {\it Gaia} (Figure \ref{fig:proper_motion_image}). The tertiaries include both WD and MS stars (Figure \ref{fig:CV_HR_Diagram}) with typical separations $10^2-10^4$~au (Figure \ref{fig:distance_sep_CVs}). Our search suggests that $\gsim 10\%$ of CVs have wide tertiaries (Figure \ref{fig:CV_triple_fraction}).
    % Among the tertiary stars, three are white dwarfs, while the rest are distributed along the main sequence (Figure \ref{fig:CV_HR_Diagram}).
    \item {\bf Theoretical Investigation of the Formation of CVs in Triples:} We evolve a population of $2000$ triples using detailed three-body simulations to investigate the different evolutionary pathways for CVs in triples (Figure \ref{fig:CVs_a1_e1_a2}). When compared to binary models, our results show that $\sim60\%$ of the inner binaries that become CVs would not have become CVs in isolation. For the remaining $\sim40\%$, while these inner binaries would have become CVs without the tertiary, their formation histories often deviate from typical isolated binary evolution and involve dynamical triple channels (Section \ref{subsec:CV_formation}). The formation pathways for CVs in triples can be summarized as follows:
    \begin{enumerate}
        \item {\it High Eccentricity Formation (no CE):} 
        With the presence of a tertiary stellar companion, we predict that $20\%$ of CVs in triple form without ever experiencing a common envelope phase. These CVs begin as wide binaries ($\sim10-1000$~au) where the white dwarf formed without its progenitor ever transferring mass to the secondary. Then, EKL oscillations excited the binary's eccentricity, allowing tides and mass transfer to shrink the orbit, which initiated Roche lobe overflow, forming the CV. This EKL channel generally favors slightly evolved or high-mass donor stars because their larger radii make tidal capture with a WD more favorable. 
        \item {\it High Eccentricity Formation (with CE)}: Before the WD forms, $60\%$ of wide CV progenitors are driven into highly eccentric orbits induced by the outer tertiary. These eccentric orbits increase the number of mass transfer interactions within the inner binary, particularly when the primary star is on the giant branch. In most cases, the binary first circularizes, after which the primary evolves further, eventually filling its Roche lobe and initiating mass transfer (Figure \ref{fig:CVs_a1_e1_a2}, middle). In other cases, the mass transfer with the RG primary begins while the orbit is highly eccentric. Compared to isolated binaries, the presence of a tertiary induces higher eccentricities, enabling more inner binaries, many of which would have never interacted without a tertiary present, to go through a common envelope stage.
        \item {\it Classical CV Channel}:
        The rest of the population ($\sim20\%$) is consistent with the usual CV formation in an isolated binary. In these cases, the gravitational influence of the tertiary was small, and the inner binary became a CV independently. 
        \end{enumerate}
    \item {\bf Predicted Properties of CV triples}: 
    By combining our simulations with observations, we make several predictions for the characteristics of the CV triple population.
        \begin{enumerate}
            \item {\it CV triples are wider than field binaries:}
            Our findings show that observed CV triples are, on average, wider than resolved {\it Gaia} binaries (Figure \ref{fig:sep_hists_dist}). The wide nature of CV triples is naturally produced in our simulations and is attributed to their dynamical formation in triples. In our triple simulations, the majority of CV progenitors began as somewhat wide ($5-1000$~au) MS+MS inner binaries that later tightened through EKL processes. This would require that their tertiaries were proportionately wider initially such that the triple remained long-term stable (see Section \ref{subsec:ICs}). Therefore, compared to regular binaries, this added constraint for CVs in triples causes an excess of tertiaries with $s>1000$~au, which is present in observations. This consistency between the observed and simulated CV triples population is depicted in Figure \ref{fig:sep_hists_dist}.
            \item {\it CVs formed through triple channels tend to have longer periods at the onset of mass transfer:} As studied here, CVs forming via triple dynamics often originate from wide progenitor binaries ($\sim1-10^3$~au) that later initiate a mass transfer via the EKL mechanism. These channels favor longer CV periods at the onset of mass transfer and typically involve more evolved or more massive main-sequence donors (Figure \ref{fig:CV_periods}, left panel). In the observed population, CVs in triples also exhibit a tendency for longer orbital periods (Figure \ref{fig:CV_periods}, right panel).
            \vspace{-0.1em}
            \item {\it Number of CV triples in the Galaxy:} Using the results from our simulated triple population, we roughly estimate that $4.7\times10^5$ CVs in the Galaxy formed in the presence of a tertiary star.
        \end{enumerate}
\end{enumerate}

\section{Acknowledgments} \label{acknowledgments}
 We thank the referee for constructive comments on the manuscript. C.S. thanks Ed Nathan, Natsuko Yamaguchi, and Samuel Whitebook for useful discussions. C.S. is supported by the Joshua and Beth Friedman Foundation Fund. S.N. acknowledges the partial support from NASA ATP 80NSSC20K0505 and from the NSF-AST 2206428 grant, as well as thanks Howard and Astrid Preston for their generous support. This research was supported by NSF grant AST-2307232.

\appendix 
\vspace{-2em}
\section{Folded ZTF light curves and Recovered Periods}\label{app:folded_LCs}

\begin{figure*}[h]
    \centering
    \begin{minipage}[t]{0.49\textwidth}
        \includegraphics[width=\textwidth]{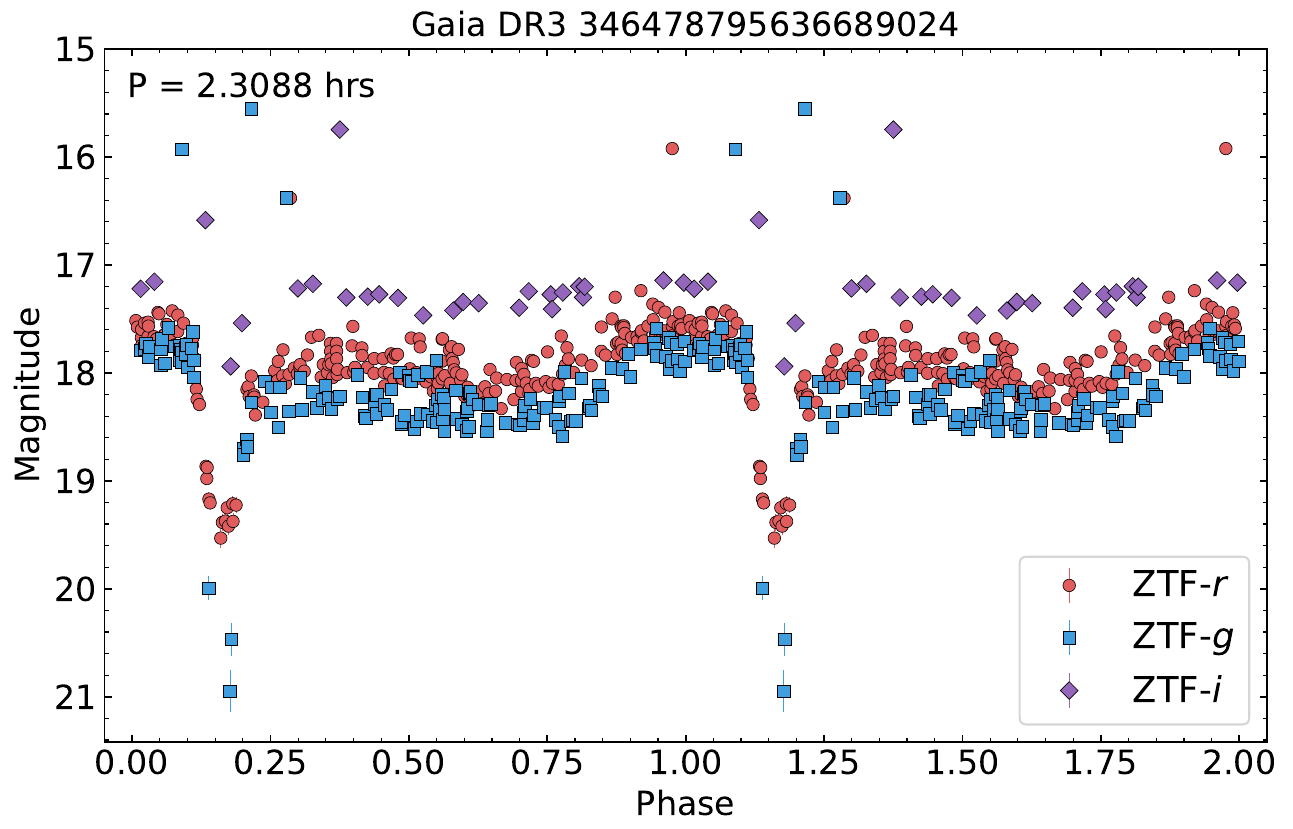}
    \end{minipage}
    \hfill
    \begin{minipage}[t]{0.49\textwidth}
        \includegraphics[width=\textwidth]{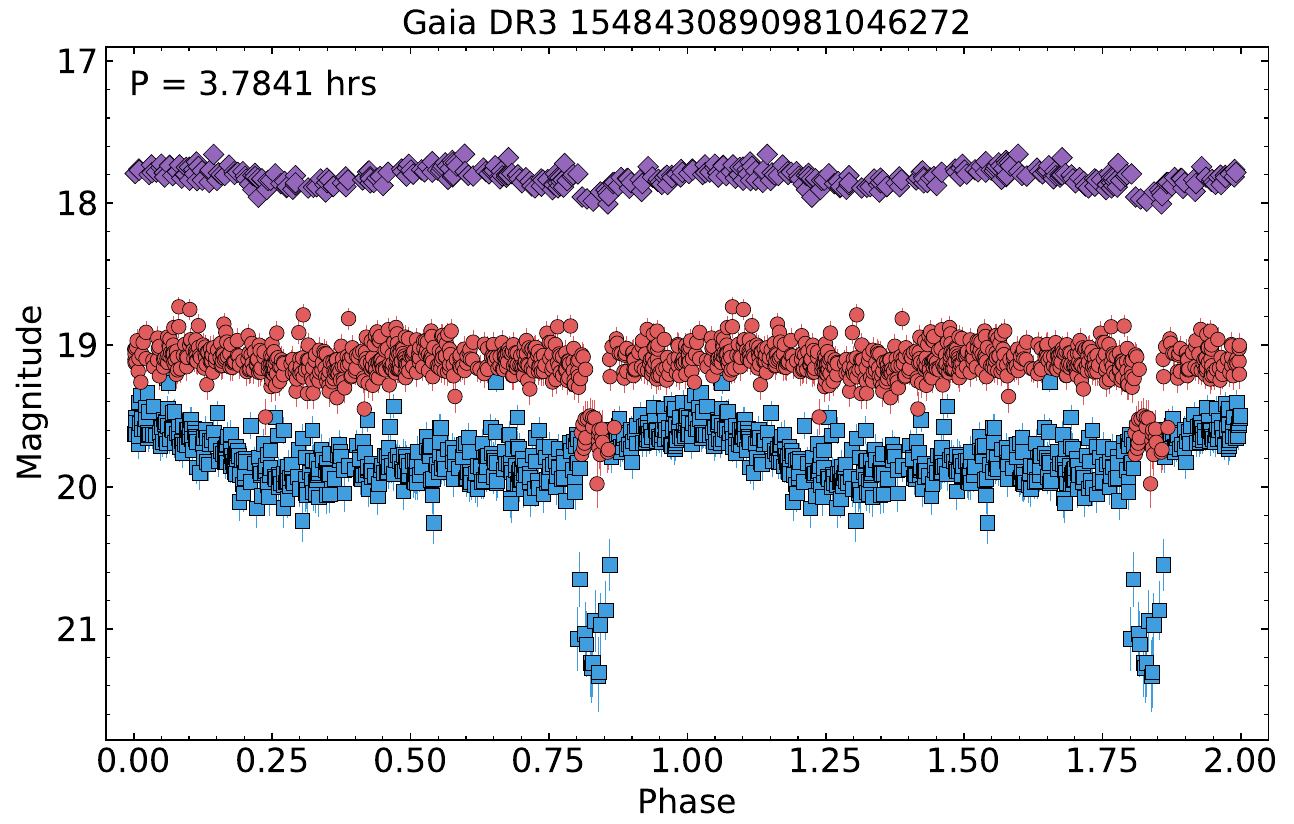}
    \end{minipage}
    \vspace{0.1cm}  % Adjust spacing between rows
    \begin{minipage}[t]{0.49\textwidth}
        \includegraphics[width=\textwidth]{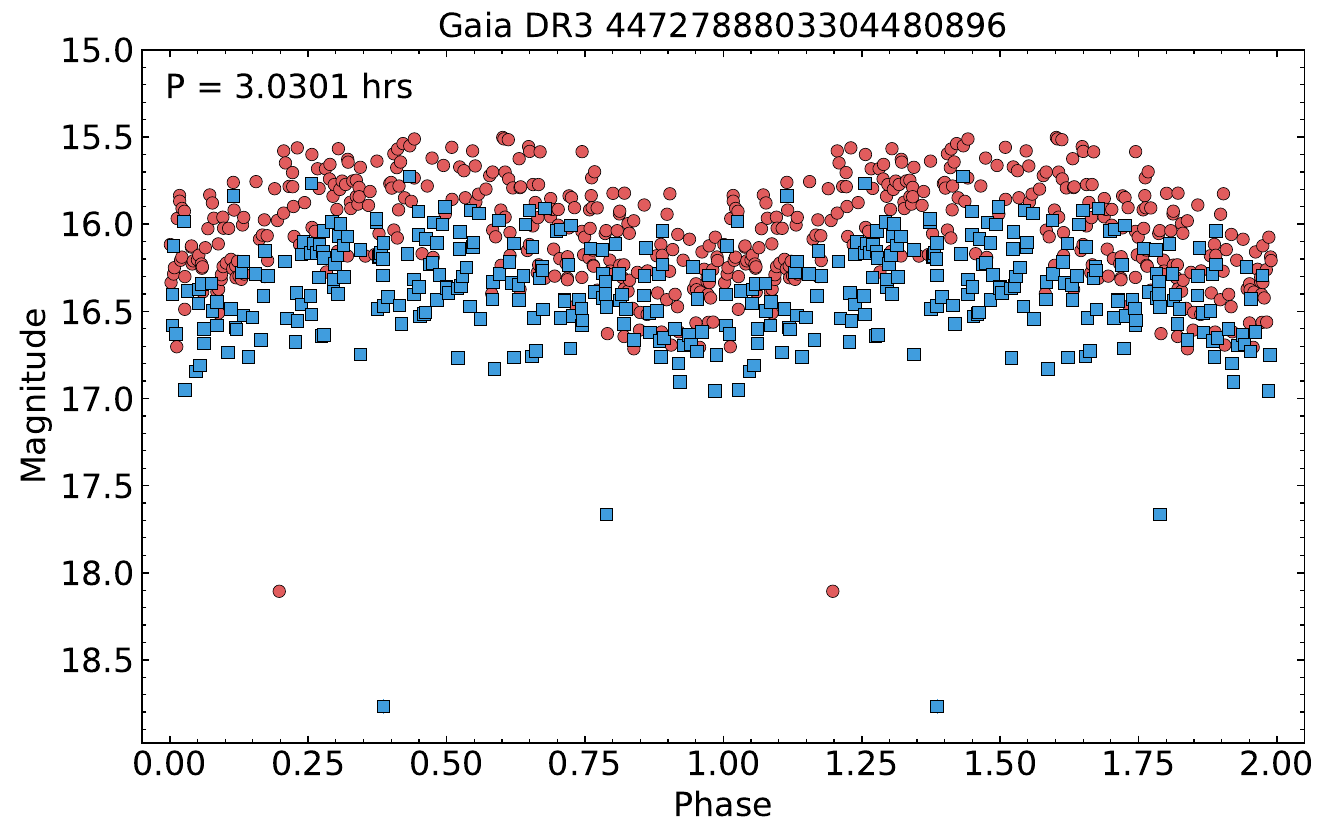}
    \end{minipage}
    \hfill
    \begin{minipage}[t]{0.49\textwidth}
        \includegraphics[width=\textwidth]{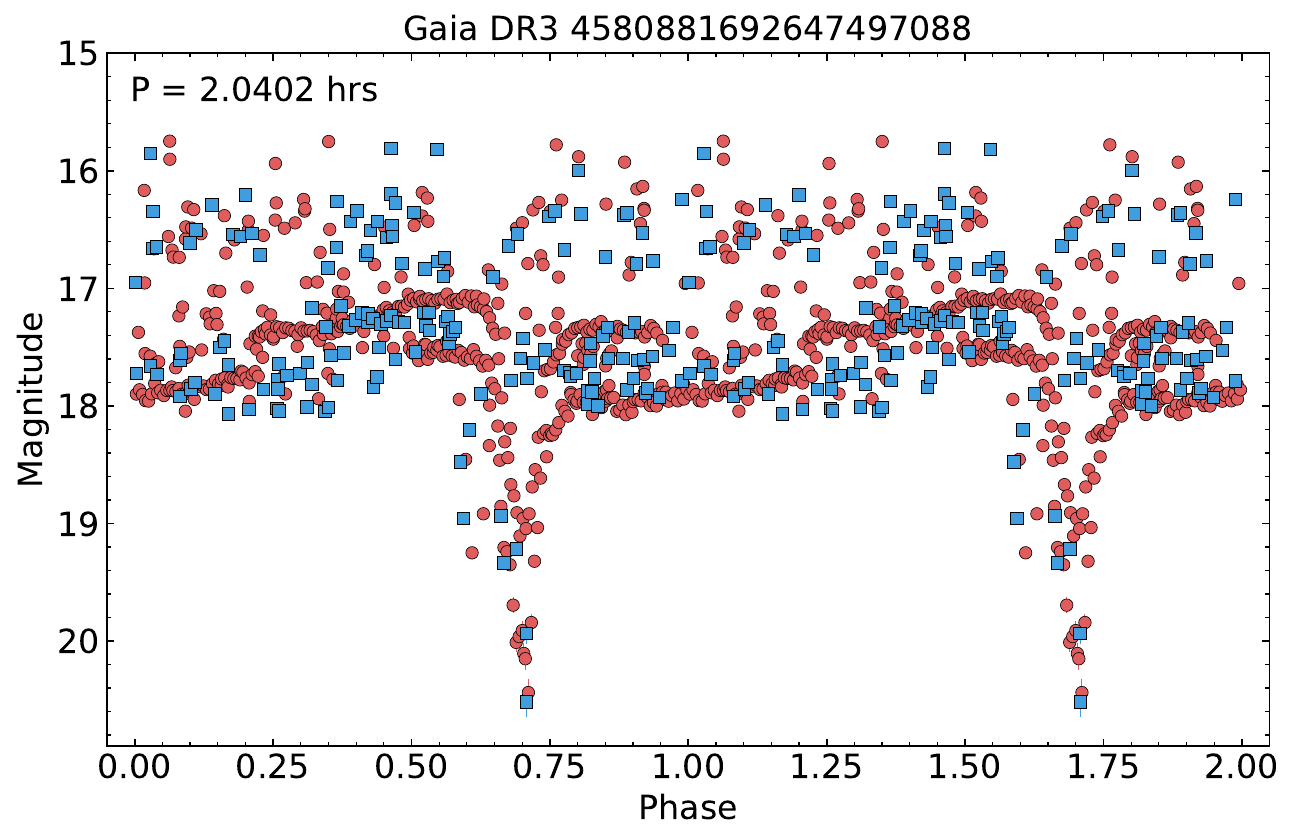}
    \end{minipage}
    \caption{Phase-folded light curves for four systems CVs without previously reported VSX periods. Each panel shows the measured ZTF $r$, $g$, and $i$ band phase-folded light curves with binned points included. The {\it Gaia} source ID along with the best-fit period are displayed for each system.}
    \label{fig:phase_folded_LCs}
\end{figure*}

For each CV in our sample, we perform a Lomb-Scargle periodicity search, and for four of them, we note robust periodicities that were not previously reported in VSX. Their folded light curves are shown in Figure \ref{fig:phase_folded_LCs}. 

Three of the CVs in Figure \ref{fig:phase_folded_LCs} are eclipsing, while the other one ({\it Gaia } DR3 4472788803304480896) shows sinusoidal variability. {\it Gaia } DR3 4472788803304480896 also recently shifted to a fainter state in its optical lightcurve, which was similarly observed near BJD=2459000 days (Figure \ref{fig:ztf_lcs}). The periods of these CVs are all less than 10 hours, with the shortest eclipsing period being $2.04$~hour ({\it Gaia }DR3 4580881692647497088). Another interesting feature is that this system's light curve seems to have $3$ different baseline flux values during different periods of quiescence. This is also seen in its raw light curve in Figure \ref{fig:ztf_lcs}, where the systems seems to exhibit variability on many timescales. The other three CVs have periods between $4$ and $8$ hours, which are consistent with being CV orbital periods.

\section{X-ray Binary or Cataclysmic Variable?}\label{app:IGR}

Two of our CV triples were classified as X-ray binaries (XRBs) in the VSX catalog. These are IGR J19308+0530, which was classified as an Intermediate-Mass X-ray Binary \citep[IMXB;][]{Fortin18} and SWIFT J2124.6+0500, which was classified as a Low-Mass X-ray Binary \citep[LMXB;][]{Halpern13}. Both of these XRB candidates have long orbital periods ($15$ and $20$ hours), and their nature as XRBs has been debated. We investigate these systems in more detail to understand whether they are wide CVs or instead harbor a neutron star. In the end, we include both of these systems in our CV sample for reasons that we discuss below. 
% However, for SWIFT J2124.6+0500, we still do not confidently rule out a neutron star primary.  

\subsection{IGR J19308+0530}
IGR J19408+0530 was first detected by the Integral satellite as an unclassified X-ray source \citep[][]{Bodaghee07}.
The proposed X-ray counterpart has been studied with both {\it Swift} and {\it Chandra}, which localized this source with sub-arcsecond resolution. Initially, \citet{Ratti13} suggest that the system is consistent with being a CV, though they note that a neutron star accretor cannot be ruled out. Later on, \citet{Fortin18} propose that the system is more consistent with having a neutron star accretor, making it an IMXB. Given that the system has a relatively close tertiary companion ($720$~au), we investigate it in more detail. 

The Swift spectrum is best-fit by a 0.2 keV blackbody with $N_H<1.5\times10^{21}$~{erg cm$^{-2}$ s$^{-1}$, which corresponds to a 2-10 keV flux of $3.3\times10^{-13}$~erg cm$^{-2}$ s$^{-1}$ \citep{Rodriguez08}. \citet{Ratti13} identify an optical counterpart of spectral type F4V and find its radial velocity (RV) to vary with a period of $0.610920$~days ($14.66$~hrs); they associate this period with the orbital period of the binary. They measure a projected rotation velocity $v\sin i =108.9 \pm 0.6 $~km s$^{-1}$ and RV semi-amplitude $K_2=91.4\pm1.4$~km s$^{-1}$.
From this, they infer a mass ratio $q = 1.78 \pm 0.04$ \citep[see Equation \ref{eq:RLO_q};][]{Horne86}. For a typical F4 star with a mass of $\sim1.4$~M$_\odot$, this would imply a faint companion mass of $\sim0.8$~M$_\odot$, which is consistent with a WD.

\begin{equation}\label{eq:RLO_q}
    \frac{v \sin i}{K_2} = (1 + q) \frac{0.49 q^{2/3}}{0.6 q^{2/3} + \ln(1 + q^{1/3})}
\end{equation}

From {\it Gaia}, we learn that the system is $331$~pc away and contains a third component $\sim 700$~au away. Given this distance, we can constrain the 2-10 keV X-ray luminosity to be $4.35\times10^{30}$~erg s$^{-1}$. {\it Gaia} also reports an RV for the tertiary of $-48.5 \pm 4.0$~km s$^{-1}$. 

From the ZTF light curve, we infer a period of $0.6109$~days, consistent with \citet{Ratti13}. We obtained 4 spectra of the source using the Fiber-fed Extended Range Optical Spectrograph \citep[FEROS;][]{Kaufer99} on the 2.2m ESO/MPG telescope at La Silla Observatory. Our observations use $1\times 1$ binning with $900$~s exposure times, yielding a spectral resolution $R \approx 50\, 000$. The data were reduced using the CERES pipeline \citep{Brahm17}, which performs bias-subtraction, flat fielding, wavelength calibration, and optimal extraction. The pipeline also measures and corrects for small shifts in the wavelength solution during the course a night via simultaneous observations of a ThAr lamp with a secondary fiber.
We derive RVs by cross-correlating the spectra with an F-star template. We then use MCMC to simultaneously constrain the following RV parameters: systemic velocity ($\gamma$), RV semi-amplitude ($K_2$), and reference time ($T_0$), assuming a circular orbit. From our four RVs, we infer $\gamma = -50.4 \pm 4.7$~km s$^{-1}$ and $K_2=95 \pm 5.9$~km s$^{-1}$. We plot the corresponding RV curve in Figure \ref{fig:IGR_RV}. Overall, our parameters are consistent with those of \citet{Ratti13}, and the $\gamma$ is consistent with the RV of the tertiary star as measured by {\it Gaia}, providing further evidence that the sources are indeed gravitationally bound. If we assume a projected rotation velocity $v\sin i=109$~km s$^{-1}$ following \citet{Ratti13}, we derive a consistent estimate for the mass ratio of $q\approx 1.7$. For a typical F4 star with $M_{\rm donor}=1.4$~M$_\odot$, this implies an accretor mass of $M_1=0.82$~M$_\odot$, consistent with a WD and with the results of \citet{Ratti13}.

\citet{Fortin18} and \citet{Avakyan23} argue that IGR J19408+0530 is likely an IMXB (i.e., that it has a neutron star accretor) based primarily on its soft X-ray spectrum. Given the companion's low inferred mass and the source's low X-ray luminosity, we consider a WD companion much more likely and therefore include it in our sample. We additionally note that while all previous work has assumed the companion must be a compact object, we see no reason to exclude the possibility that it could be another luminous star and that the X-rays could be a result of coronal activity, making the source an RS CVn binary. Future observations may clarify the nature of the accretor.

%All of the above papers regarding the nature of IGR J19408+0530 assume that the secondary is a compact object, and therefore the system must either be a CV or XRB. However, there is not concrete observational evidence that the system contains a compact object at all. In the case that it is a MS star, the system would be a RS Canum Venaticorum-type  (RS CVn) system. The ratio of the X-ray flux to the optical flux (from {\it Gaia}) also is similar with other RS CVn active binaries \citep[][]{RodriguezMS}. 

% based on its spectral analysis and K-band spectroscopy. Specifically, \citet{Fortin18} note a prominent $\text{Br}\gamma$ absorption line and weaker absorption lines from NaI and CaI. From these, they conclude that the star is consistent with an intermediate mass MS star, and along with the X-ray flux, consistent with an IMXB. However, they never determine whether there exists a neutron star in the system.
%Future follow-up observations may make the nature of the accretor in this system more clear, but for the purposes of this analysis, we incorporate it into our CV sample. Also, note that since the tertiary star is relatively close to the inner binary, its gravitational influence on this will be undoubtedly significant over long timescales. Regardless of the nature of the primary, the system is a strong candidate for having formed initially wide and being brought into contact at a later stage through three-body dynamical evolution.

\subsection{SWIFT J2124.6+0500}
SWIFT J2124.6+0500 is another system for which the nature of the accretor is unclear. The orbital period of the inner binary, $0.833$~days \citep[$20$~hrs;]{Halpern15, Bruch24}, is relatively long for a CV. We find that the system harbors a bound companion at a separation of $5300$~au, and the {\it Gaia} parallax implies a distance of $880$~pc. We derive a 2-10 keV X-ray luminosity of $L_X=1.12\times10^{33}$ erg s$^{-1}$, which is relatively high compared to most CVs \citep[e.g.,][]{Rodriguez24}. The optical spectrum shows broadened emission lines of H, He I, HeII, and CIII/NIII, which are consistent with a high accretion rate. The system is located at a Galactic latitude of $-30.5^\circ$, and based on its emission features and X-ray luminosity, was initially classified as an LMXB \citep{Halpern13}. Later on, \citet{Parisi14} and \citet{Halpern15} classified the system as a nova-like variable (i.e., a CV with a high accretion rate and a disk permanently in the high state) on the basis of its optical spectrum.
% claim that the low He II/H$_\beta$ ratio implies a non-magnetic nature, making it more consistent with a CV.
The existence of a bound wide companion suggests that large kicks were absent during the evolution of this system. We consider a WD companion the most likely scenario. 

\begin{figure}[h]
    \centering
    \includegraphics[width=0.75
    \textwidth]{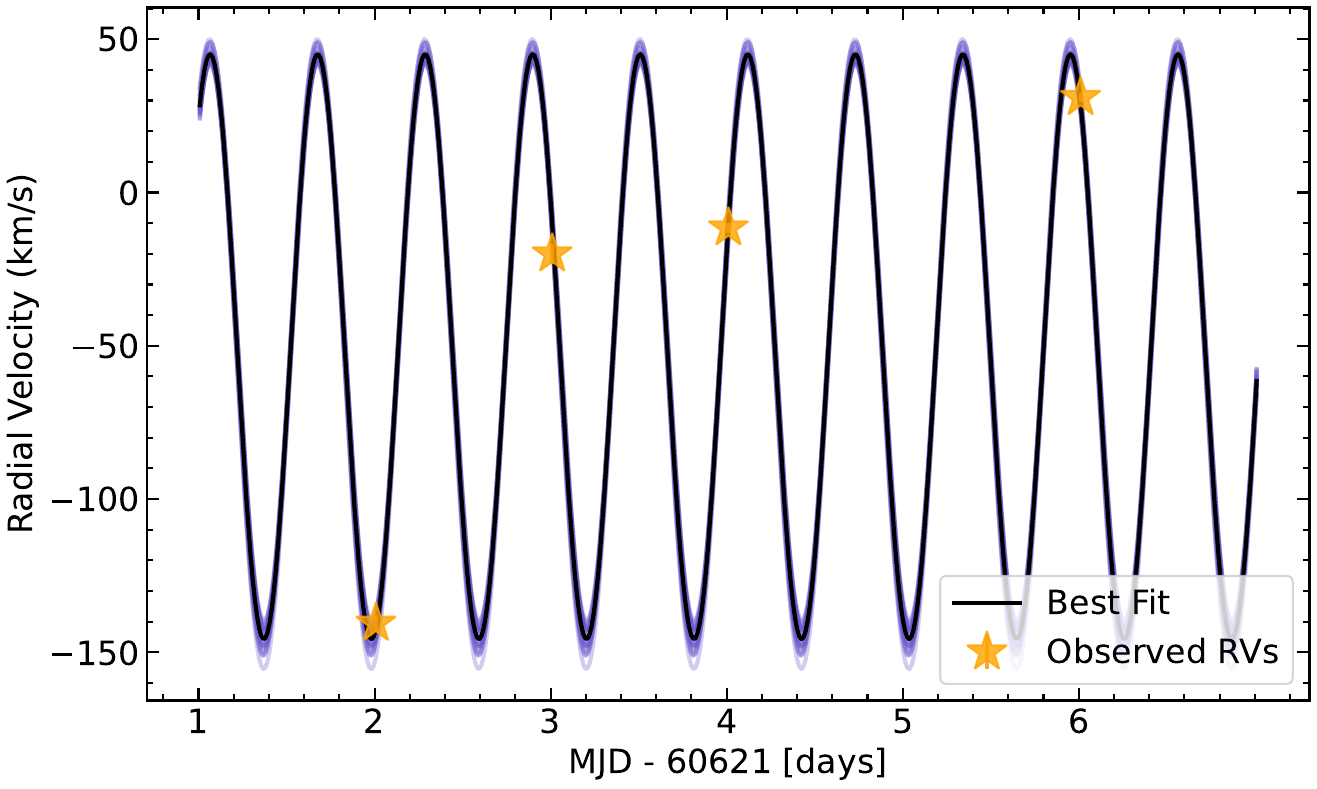}
    \caption{Radial velocity curve for IGR J19308+0530. In orange, we show the measured RVs and their uncertainties. In black we show our best-fit circular orbit folded on a period of $14.6$ hours. The purple lines show predictions for random samples from the posterior of our RV fit.
    }\label{fig:IGR_RV} 
\end{figure}

\section{Magnetic Braking}\label{app:MB}
We modify {\tt COSMIC} to include the MB prescription from \citet{Rappaport83},  which depletes the angular momentum of the binary through the following torque: 
\begin{equation}\label{eq:mb_rappaport}
    \tau_{\rm mb} = -6.8\times 10^{-34} \left(\frac{M_{2}}{M_\odot}\right) \left(\frac{R_{2}}{R_\odot}\right)^{\gamma_{\rm mb}} \left(\frac{P_{\rm orb}}{{\rm d}}\right)^{-3} \text{dyn cm},
\end{equation}
where subscript `2' denotes the donor star.
This expression assumes that the binary is tidally locked and that the donor's angular momentum loss through winds is the same as for an isolated single star with the same rotational velocity \citep{Skumanich72,Verbunt81}. 
We adopt $\gamma_{\rm BM}=3$ and switch off MB when the donor becomes fully convective. This MB law has been shown to produce CV populations broadly consistent with observations, though the precise functional form remains uncertain \citep[e.g.,][]{Andronov03, Rappaport83,EB21CV, EB22MB}.

The default MB recipe in {\tt COSMIC} has a similar form but uses the mass of the donor's envelope instead of its mass \citep[see Equation 50 in][]{BSE}.
As shown in Figure \ref{fig:MESA_vs_COSMIC_MB}, the updated MB model leads to consistent evolution between the {\tt COSMIC} CVs and CV evolution in {\tt MESA} until near the upper edge of the period gap.

\begin{figure}
    \centering
    \includegraphics[width=0.49
    \textwidth]{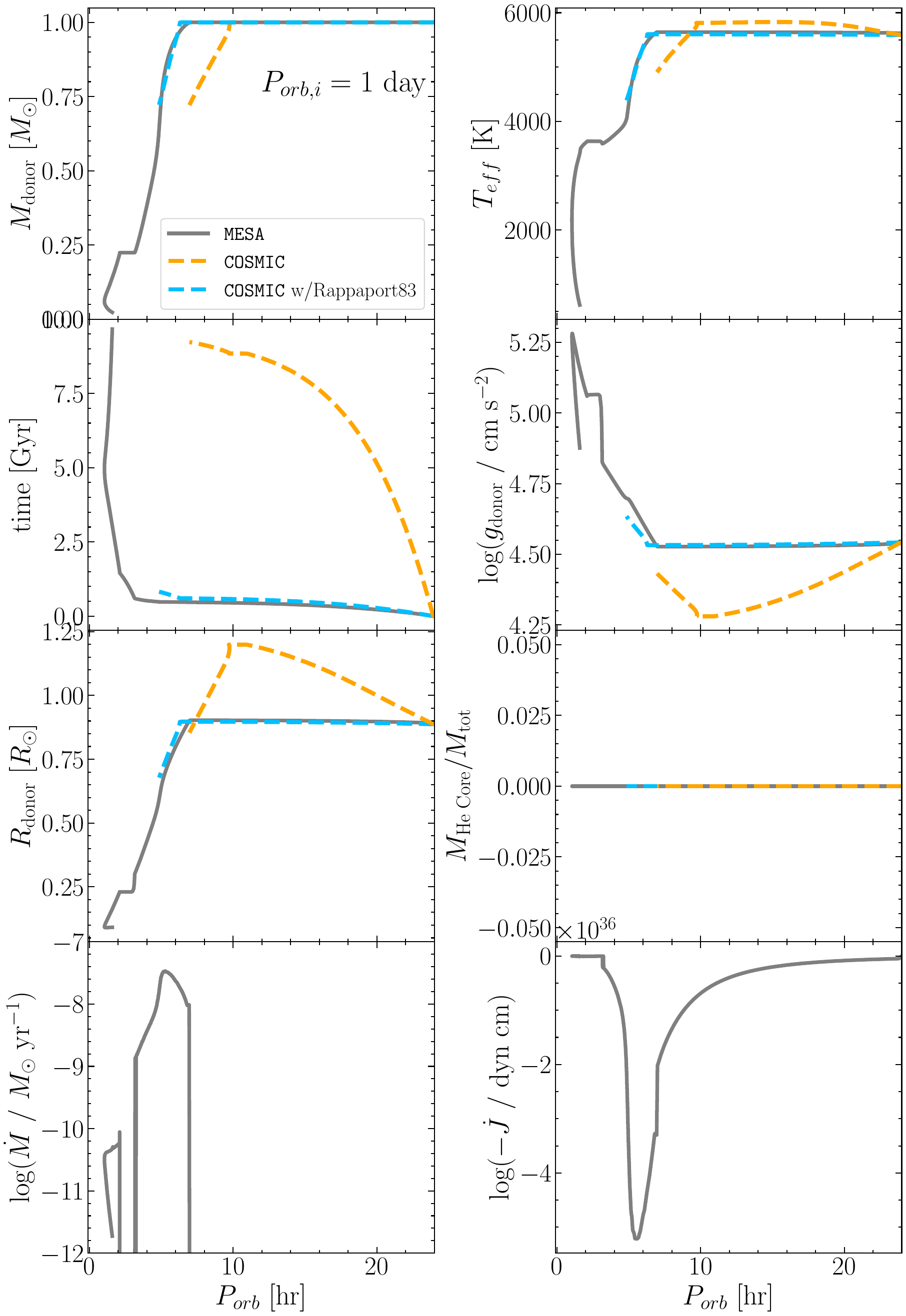}
    \caption{Comparison between {\tt MESA} and {\tt COSMIC} binary evolution using different magnetic braking prescriptions. All tracks show the evolution of a WD+MS binary with $M_{\rm \text{WD}}=0.9$~M$_\odot$, $M_{\rm \text{donor}}=1$~M$_\odot$, and $P_{\rm \text{orb}}=1$~day initially. The black curve shows the evolution in {\tt MESA}, and the orange curve shows the evolution in {\tt COSMIC} with its default magnetic braking prescription. In green, we plot  {\tt COSMIC} evolution after the magnetic braking prescription as changed to match that of {\tt MESA}, which now both follow \citet{Rappaport83}.}\label{fig:MESA_vs_COSMIC_MB} 
\end{figure}

To assess the effects of a different magnetic braking model (Equation \ref{eq:mb_rappaport}), we run a set of binary {\tt MESA} models and compare them to {\tt COSMIC}. Using {\tt MESA} version {\tt r24.03.1}, we evolve a WD+MS binary with $M_{\rm \text{WD}}=0.9$~M$_\odot$ and a donor with mass varying from $M_{\rm \text{donor}}=1-3$~M$_\odot$ in steps of $0.5$ and initial orbital period ranging from of $1-3$~days in steps of $0.5$. We assume solar metallically and a point-mass WD. We evolve the same binaries with identical initial conditions in {\tt COSMIC} and compare the outputs in Figure \ref{fig:MESA_vs_COSMIC_MB}. In this Figure, we plot the {\tt MESA} binary evolution and {\tt COSMIC} with default (orange) and updated (blue) MB recipes. Here, we show the results for a binary that is initialized with $M_{\rm \text{donor}}=1$~M$_\odot$ and $P_{\rm \text{orb}}=1$~day. 

The updated {\tt COSMIC} code is consistent with {\tt MESA} until the orbital period reaches $\approx 4.5$~hrs, which is expected given that we modify {\tt COSMIC} to use the same MB model. The default MB implementation in {\tt COSMIC}, on the other hand, results in much weaker MB, such that mass transfer only begins several Gyr later. Even with our updated MB prescription {\tt COSMIC} merges the binary at an apparently arbitrary time, when the orbital period is $\sim4.5$~hrs, because it does not follow the evolution of the donor's stellar structure in detail. With the new magnetic braking, this occurs after the binary reaches an orbital period of $5$~hrs, but with the default MB prescription in COSMIC, this occurs when the binary has $P_{\rm \text{orb}}\approx20$~hrs, shortly after the onset of mass transfer (second row of Figure \ref{fig:MESA_vs_COSMIC_MB}). Such premature mergers reduce the amount of time that a stably mass-transferring WD+MS binary lives.

% One key feature of the default magnetic braking prescription in {\tt COSMIC} \citep[from equation 50 of][]{BSE} is such the binaries merge rapidly after mass transfer, as is seen for the orange curve in the second row of Figure \ref{fig:MESA_vs_COSMIC_MB}. In contrast, our updated MB prescription thus leads to CVs that remain as stably mass-transferring CVs for a longer period of time (blue curve), as is consistent with the long-lived lifetimes of CVs \citep[e.g.,][]{Rappaport82}.
Given the greater consistency with {\tt MESA} (Figure \ref{fig:MESA_vs_COSMIC_MB}), we adopt the \citet{Rappaport83} MB prescription to follow the evolution of our inner binaries in {\tt COSMIC} and better model CV formation in triples.

\section{Other WD accreting inner binaries}\label{app:WD_accreting}
In our simulated triple population, most ($\sim70\%$) of the WD accreting inner binaries are not CVs and instead have somewhat evolved donor stars. These donors include He MS stars, RG stars, and Asymptotic Giant Branch (AGB) stars. In many ways, their evolution is similar to the CVs, yet their observable properties are quite different \citep[e.g.,][]{Belczynski00, Solheim10}. In Figure \ref{fig:CVs_a1_e1_a2_MT2}, we plot the characteristics of these systems (1) at the onset of the first mass transfer within our triple code and (2) at the time when they begin WD mass transfer after the inner binary was transported into {\tt COSMIC}. 
% This transfer between codes is done such that the mass transfer evolution can be followed, assuming that the tertiary plays a minimal role, which is accurate given that EKL is mostly suppressed during this stage.  
Note also that most rapid binary population synthesis codes, including {\tt COSMIC}, assume instantaneous circularization upon mass transfer, which is why all inner binaries have $e_1\sim0$ in the second panel. Note also that in this second panel, all of the primary stars are WDs, while in the first panel this is not necessarily the case. 

Figure \ref{fig:CVs_a1_e1_a2_MT2} suggests that our triple simulations can produce an array of WD accreting systems, such as WD+MS (CVs), WD+Helium MS (AM CVn), WD+RG, and WD+AGB systems. In fact, WD+RG systems are produced with the greatest abundance and have tertiary companions from $10-10^5$~au. Similarly, we find several AGB stars that transfer mass even out to $10$~au. Most of the inner binaries in Figure \ref{fig:CVs_a1_e1_a2_MT2} experience strong EKL oscillations, and thereby high eccentricities. We conclude that tertiaries are also likely to play a role in the formation of some symbiotic and AM CVn binaries; we defer a detailed study of these populations to future work.

\begin{figure*}
    \centering
    \includegraphics[width=0.75
    \textwidth]{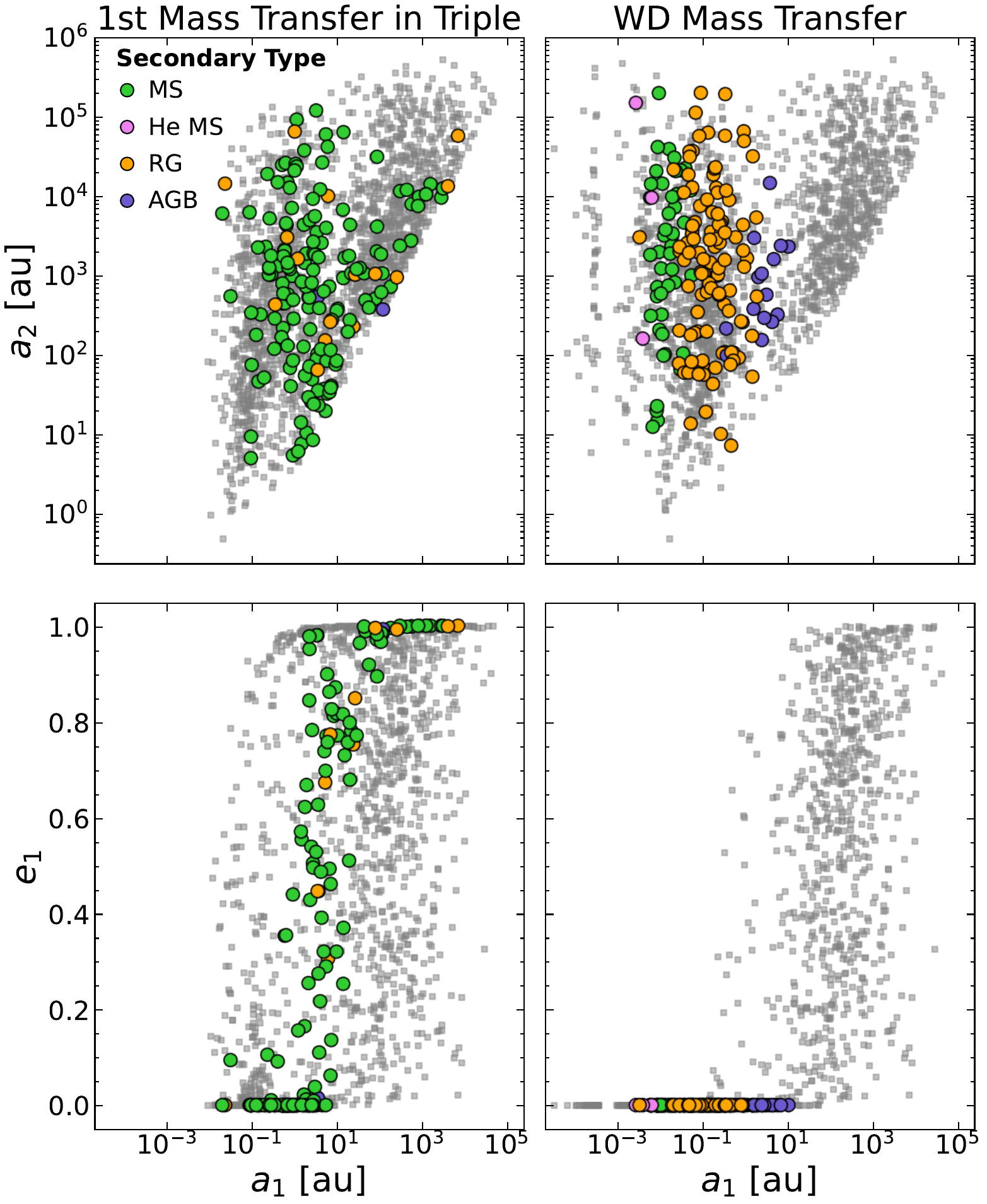}
    \caption{Orbital parameters of WD accreting systems in triples at different stages of evolution. All of the large circular points evolved to host inner binaries that contain a WD accreting from a MS or evolved companion. The first panel shows the orbital parameters at the first time of mass transfer in the triple code. The second panel shows the orbital parameters during the WD accretion stage. All systems are circularized in the second panel because the inner binaries were transferred into {\tt COSMIC} to track the mass transfer evolution, and {\tt COSMIC} assumes instant circularization at mass transfer. The color of the points represent the type of the secondary star during these mass transfer stages. In the first panel, the primary can be an MS, RG, or WD and only the secondary type is specified by the colors. In the second panel, all primaries are WDs, so the secondary type is the type of the donor star in these systems. Note that the green points in this panel are the CVs in our population (i.e., WD+MS accreting binaries). The gray background points show all triples from the simulated population, including those that did not end up as WD mass transferring systems. }\label{fig:CVs_a1_e1_a2_MT2} 
\end{figure*}

\section{Impulsive Mass Loss}\label{app:impulsive_ML}
Throughout our study, we assume that mass loss in the inner binary leads to an adiabatic widening of the outer binary. While this has been generally accepted for slow, isotropic mass loss, the physical picture may change when the mass loss is sudden and anisotropic. The two major points in CV formation where this may occur are (1) when the AGB primary becomes a WD or (2) when the binary undergoes common envelope evolution, if it does. For the first case, previous studies suggest that WD binary and triple population show imprints of WD formation kicks \citep[e.g.,][]{Fregeau09, EB18, Shariat23, Shariat24}. For rapid CE mass loss, the spiral-in phase may eject mass on timescales shorter than the tertiary star's orbital period, potentially causing an impulsive kick \citep[e.g.,][]{Toonen17, Michaely2019b, Igoshev20}. 

We test the impact of impulsive mass loss on the tertiary separation, without considering any additional kick velocity ($v_k=0$). Following the methods of \citet{Lu2019}, we calculate the change in the outer orbit's semi-major axis assuming ($v_k=u_k=0$); we plot the results in Figure \ref{fig:sep_kicks}. Similar to what was done in Figure \ref{fig:sep_hists_dist}, we sample distances for our simulated triples and only keep those that would be resolved assuming a $1''$ angular resolution. This distribution is wider than those without kicks (Figure \ref{fig:sep_hists_dist}), as would be expected. While this is more discrepant with the observed {\it Gaia} CV triples than the no-kick simulations, the low sample sizes in each sample make this discrepancy difficult to assess statistically.

\begin{figure}[h]
    \centering
    \includegraphics[width=0.75
    \textwidth]{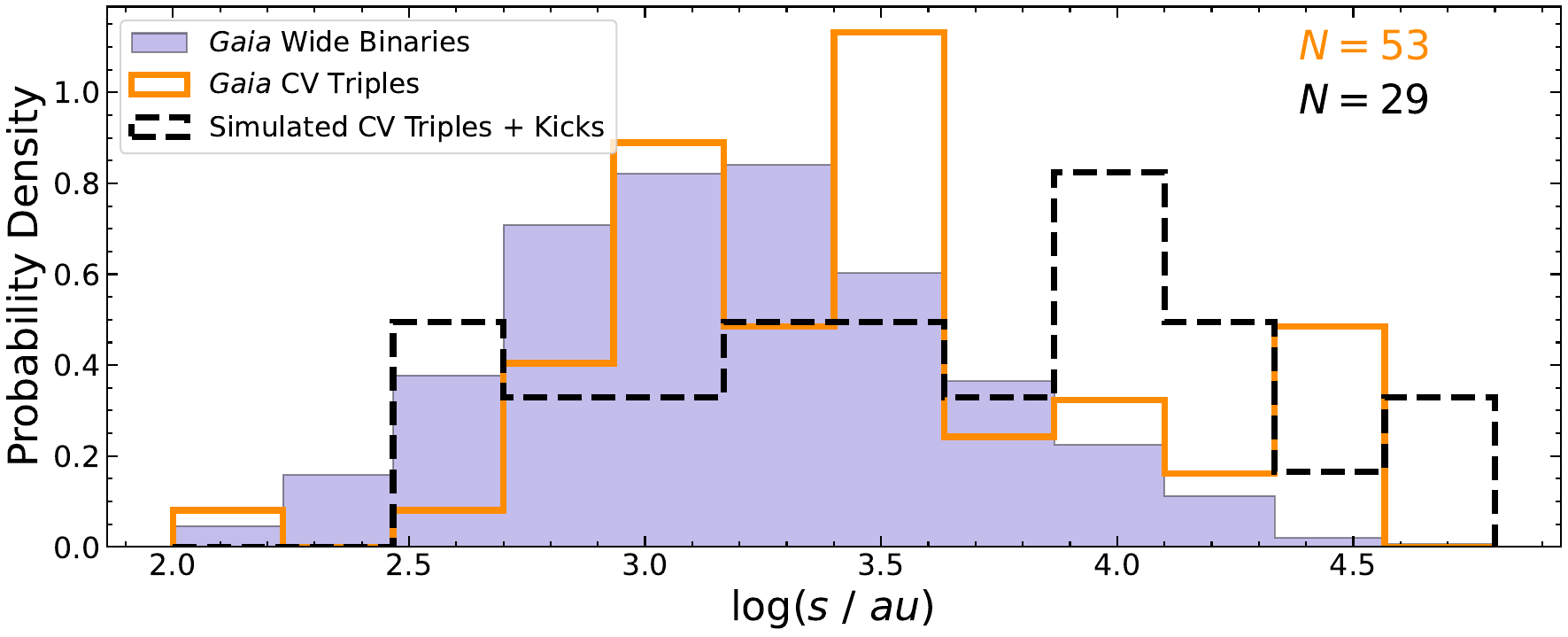}
    \caption{Same as Figure \ref{fig:sep_hists_dist}, but here we consider the effect of impulsive mass loss for the simulated CV triples (black curve).} \label{fig:sep_kicks} 
\end{figure}

\section{ZTF light curves and full tables} \label{app:data}
In Figure \ref{fig:ztf_lcs} and Table \ref{tab:CV_gaia_params_app} we respectively show the ZTF light curves (when available) and {\it Gaia} parameters for our sample. 

\begin{figure*}
\centering
\includegraphics[width=0.90\textwidth]{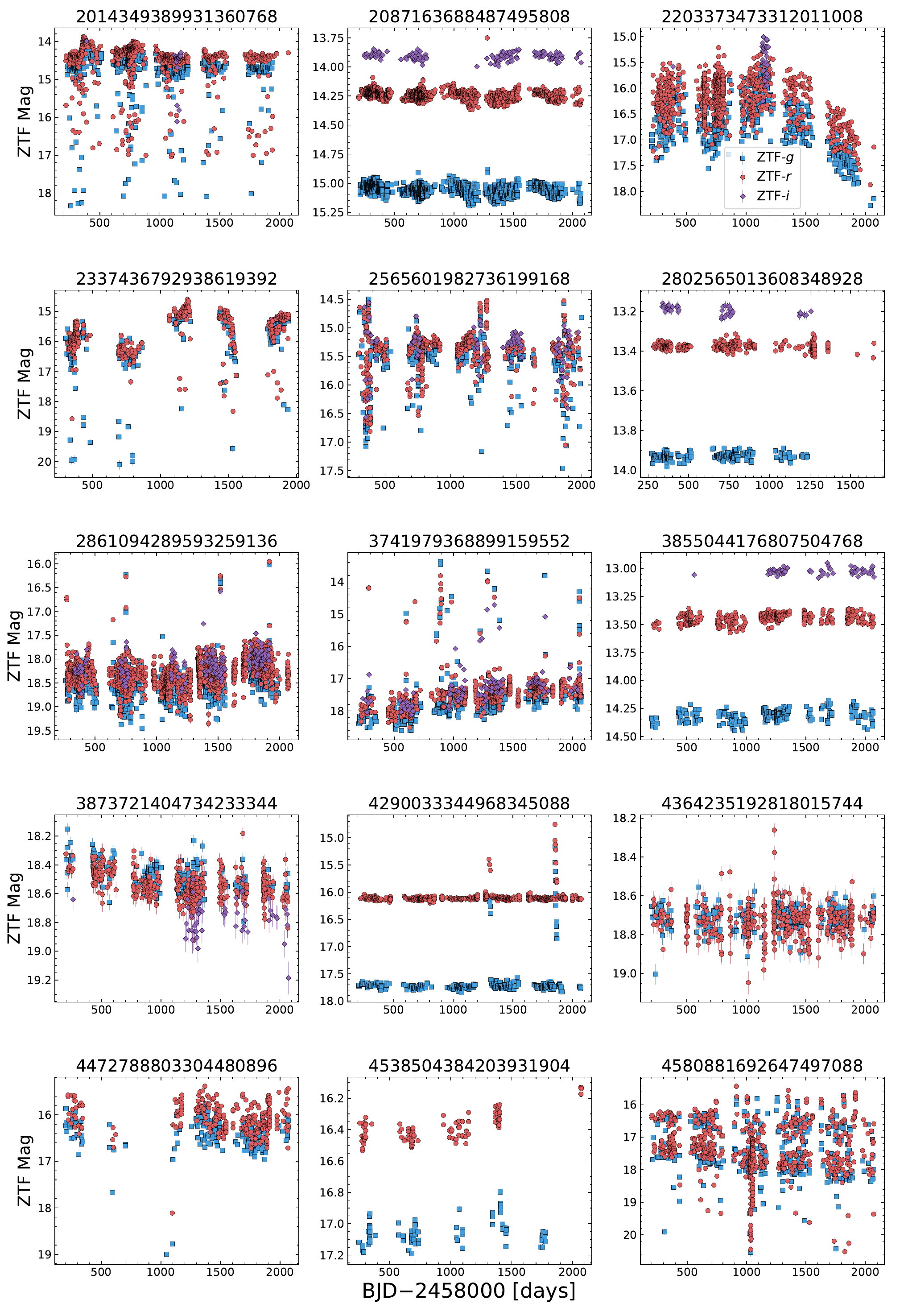}
\caption{Optical light curves for all CV triples with ZTF data. We show available photometry from the ZTF g, r, and i band and label each plot by the {\it Gaia} DR3 source id. }\label{fig:ztf_lcs}
\end{figure*}

\begin{figure*}
% \ContinuedFloat
\centering
\includegraphics[width=0.90\textwidth]{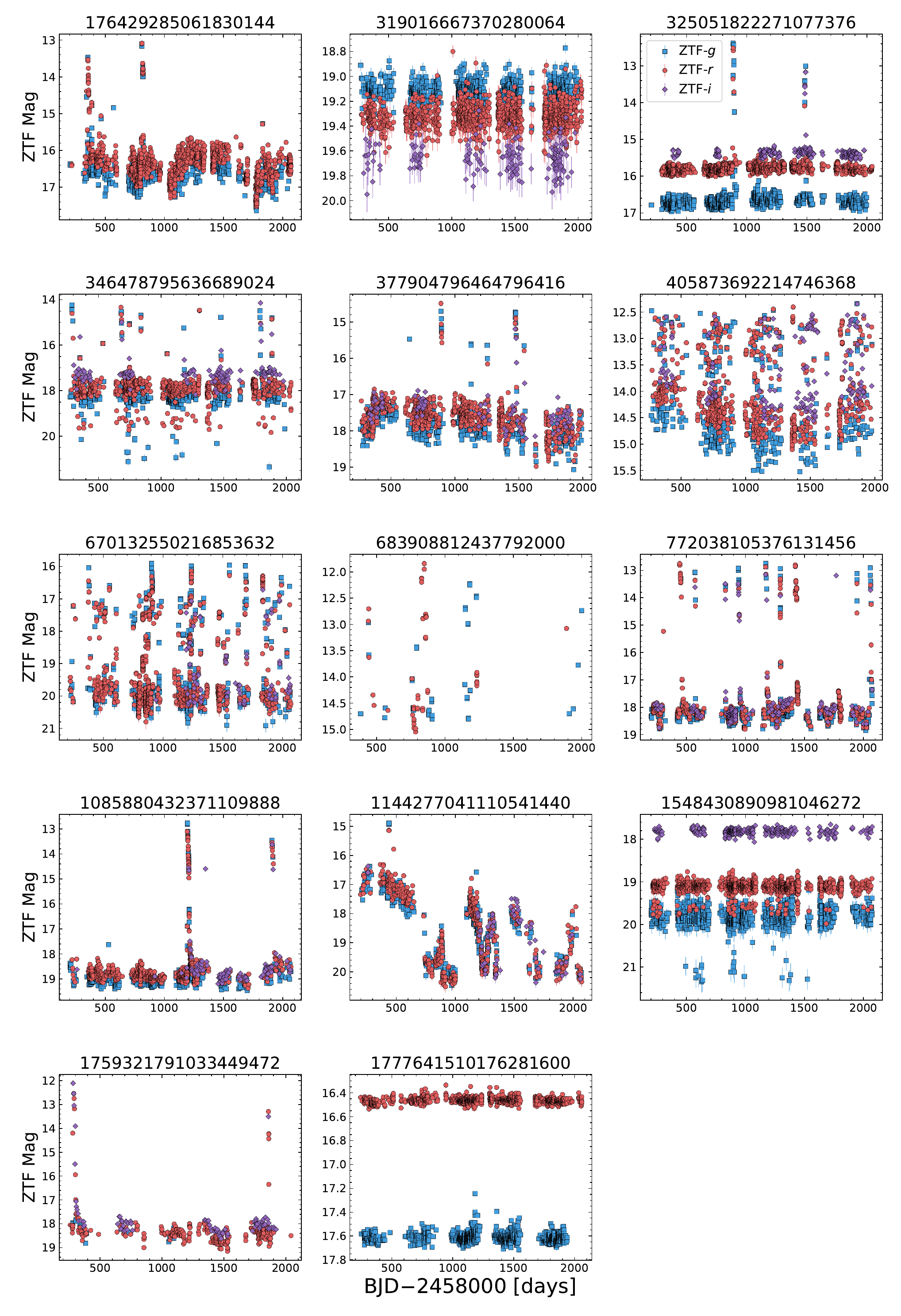}
\caption{Continuation of Figure \ref{fig:ztf_lcs}.}
\end{figure*}

\begin{longrotatetable}
\begin{deluxetable}{lccccccccc}
\centerwidetable
\tablecaption{{\it Gaia} parameters of CVs with Wide Tertiary Companions\label{tab:CV_gaia_params_app}. Subscript `1' refers to the CV and `2' refers to the wide companion.}
\tablehead{
\colhead{Name} &
\colhead{{\it Gaia }DR3 ID 1} &
\colhead{{\it Gaia }DR3 ID 2} &  
\colhead{$\varpi_1$} & 
\colhead{$\varpi_2$} &
\colhead{$(G_{\rm BP}-G_{\rm RP})_1$} &
\colhead{$(G_{\rm BP}-G_{\rm RP})_2$} &
\colhead{$M_{\rm G,~1}$} &
\colhead{$M_{\rm G,~2}$} &
\colhead{$s$} 
\\
\colhead{} &
\colhead{} &
\colhead{} &  
\colhead{[mas]} & 
\colhead{[mas]} &
\colhead{[mag]} &
\colhead{[mag]} &
\colhead{[mag]} &
\colhead{[mag]} &
\colhead{[au]}
% \colhead{[days]} 
}
\startdata
Gaia J051903.9+630339.6       & 285957277597658240  & 285957277597658368  & $8.585 \pm 0.025$    & $8.612 \pm 0.013$   & $0.569$   & $0.740$  & $9.642$            & $4.284$           & $793.1$   \\
V3885 Sgr                     & 6688624794231054976 & 6688624794233492864 & $7.753 \pm 0.034$    & $7.699 \pm 0.094$   & $0.108$   & $2.226$  & $4.652$            & $9.939$           & $890.6$   \\
V379 Tel                      & 6658737220627065984 & 6658737388128701184 & $7.714 \pm 0.057$    & $7.548 \pm 0.033$   & $1.254$   & $2.356$  & $10.610$           & $9.414$           & $4083.1$  \\
V1108 Her                     & 4538504384210935424  & 4538504384203931904 & $6.475 \pm 0.135$    & $6.758 \pm 0.100$   & $2.033$   & --       & $10.392$           & $11.553$          & $155.7$   \\
MASTER OT J042609.34+354144.8 & 176429285061830144  & 176429323716565504  & $5.375 \pm 0.050$    & $5.564 \pm 0.070$   & $0.500$   & $2.557$  & $9.307$            & $9.803$           & $3089.8$  \\
SDSS J101421.55+063857.7      & 3873721404734233344 & 3873721409029428352 & $5.281 \pm 0.193$    & $5.352 \pm 0.024$   & $0.126$   & $1.800$  & $11.781$           & $7.604$           & $1654.7$  \\
LS IV -08 3                   & 4339398736975240192 & 4339398736974078464 & $4.673 \pm 0.044$    & $3.722 \pm 0.645$   & $0.023$   & $0.005$  & $3.985$            & $12.042$          & $989.5$   \\
ATO J145.8742+09.1629         & 3855044176807504768 & 3855044172512953472 & $4.330 \pm 0.018$    & $4.149 \pm 0.097$   & $1.343$   & $2.415$  & $6.614$            & $9.947$           & $1178.8$  \\
YZ Cnc                        & 683908812437792000  & 683908808140984832  & $4.245 \pm 0.033$    & $4.526 \pm 0.360$   & $0.573$   & $0.127$  & $7.131$            & $12.502$          & $5149.8$  \\
USNO-A2.0 0825-10062737       & 4364235192818015744 & 4364235192815722240 & $4.125 \pm 0.259$    & $4.176 \pm 0.570$   & $-0.249$  & --       & $11.309$           & $12.090$          & $317.2$   \\
KT Per                        & 405873692214746368  & 405873687918663424  & $4.059 \pm 0.034$    & $4.126 \pm 0.025$   & $0.742$   & $1.539$  & $6.757$            & $7.290$           & $3838.0$  \\
EF Peg                        & 1759321791033449472 & 1759321722313141760 & $3.462 \pm 0.233$    & $3.382 \pm 0.014$   & $0.783$   & $0.932$  & $10.849$           & $4.940$           & $1600.4$  \\
UW Pic                        & 4798833587650467200 & 4798833587648969088 & $3.422 \pm 0.027$    & $3.850 \pm 0.216$   & $0.825$   & --       & $8.503$            & $12.148$          & $1502.3$  \\
CSS 151208:020401+434133      & 346478795636689024  & 346478864356164608  & $3.358 \pm 0.104$    & $3.307 \pm 0.025$   & $1.050$   & $1.116$  & $10.392$           & $6.282$           & $3756.7$  \\
SDSS J120615.73+510047.0      & 1548430890981046272 & 1548430886684193152 & $3.299 \pm 0.165$    & $3.308 \pm 0.568$   & $2.195$   & $0.452$  & $11.210$           & $12.998$          & $2034.3$  \\
ASASSN-19dn                   & 5914778653652430464 & 5914778653652431104 & $3.266 \pm 0.078$    & $3.372 \pm 0.136$   & $2.710$   & --       & $9.800$            & $10.547$          & $591.2$   \\
RX J0131.4+3602               & 319016667370280064  & 319040066352107904  & $3.263 \pm 0.305$    & $3.118 \pm 0.108$   & $-0.125$  & $2.402$  & $11.559$           & $9.924$           & $8663.0$  \\
ASASSN-21cw                   & 4290033344968345088 & 4290033344954951936 & $3.192 \pm 0.042$    & $2.970 \pm 0.334$   & $1.874$   & $2.609$  & $7.919$            & $10.894$          & $703.6$   \\
RX J0154.0-5947               & 4714563374364671872 & 4714563168206242048 & $3.076 \pm 0.027$    & $3.123 \pm 0.027$   & $0.338$   & $1.846$  & $8.156$            & $8.168$           & $2728.3$  \\
MR UMa                        & 772038105376131456  & 772038105376626432  & $3.037 \pm 0.107$    & $2.735 \pm 0.035$   & $0.547$   & $1.646$  & $10.309$           & $7.517$           & $1995.7$  \\
IGR J19308+0530               & 4294249387962232576 & 4294249387935557888 & $3.013 \pm 0.027$    & $3.151 \pm 0.060$   & $0.601$   & $0.942$  & $2.566$            & $6.819$           & $722.1$   \\
Gaia DR3 2802565013608348928  & 2802565013608348928 & 2802565017902912000 & $2.406 \pm 0.018$    & $2.108 \pm 0.141$   & $0.905$   & --       & $5.238$            & $8.499$           & $519.5$   \\
DDE 202                       & 4472788803304480896 & 4472789013757913856 & $2.404 \pm 0.063$    & $2.446 \pm 0.014$   & $0.483$   & $0.735$  & $7.753$            & $3.833$           & $2573.3$  \\
MASTER OT J072948.66+593824.4 & 1085880432371109888 & 1085880436667270784 & $2.361 \pm 0.269$    & $2.638 \pm 0.040$   & $0.558$   & $1.688$  & $10.553$           & $7.663$           & $4231.4$  \\
DDE 174                       & 2203373473312011008 & 2203373477617132288 & $2.329 \pm 0.083$    & $2.263 \pm 0.013$   & $1.432$   & $0.942$  & $9.675$            & $5.085$           & $11370.3$ \\
HW Boo                        & 3741979368899159552 & 3741979815575758976 & $2.291 \pm 0.135$    & $2.232 \pm 0.165$   & $0.563$   & $2.540$  & $9.712$            & $9.775$           & $26296.4$ \\
DDE 169                       & 377904796464796416  & 377904796464796928  & $2.268 \pm 0.097$    & $1.926 \pm 0.106$   & $0.549$   & $1.253$  & $9.081$            & $5.640$           & $12985.3$ \\
ASASSN-14hw                   & 5549916268416336512 & 5549916268416335616 & $2.253 \pm 0.117$    & $2.259 \pm 0.058$   & $0.516$   & $2.168$  & $10.338$           & $8.947$           & $25469.4$ \\
DDE 202                       & 4580881692647497088 & 4580881692647172480 & $2.174 \pm 0.057$    & $2.148 \pm 0.013$   & $0.493$   & $0.848$  & $8.903$            & $4.404$           & $3548.2$  \\
J2256+5954                    & 2014349389931360768 & 2014349389931359616 & $2.065 \pm 0.017$    & $2.091 \pm 0.013$   & $0.420$   & $0.964$  & $5.541$            & $4.826$           & $2825.6$  \\
AH Men                        & 5207384891323130368 & 5207385651533430912 & $2.043 \pm 0.014$    & $2.005 \pm 0.012$   & $0.080$   & $0.830$  & $4.724$            & $4.988$           & $1439.1$  \\
HS 0218+3229                  & 325051822271077376  & 325051817976249600  & $2.018 \pm 0.042$    & $1.923 \pm 0.247$   & $1.272$   & $2.497$  & $7.190$            & $10.301$          & $2801.9$  \\
ZTF18abmneqb                  & 2861094289593259136 & 2861094289592739328 & $1.946 \pm 0.173$    & $1.593 \pm 0.016$   & $0.531$   & $0.734$  & $9.592$            & $3.997$           & $5775.1$  \\
ASASSN-19pz                   & 5391459325544320640 & 5391459325548009984 & $1.938 \pm 0.234$    & $1.694 \pm 0.052$   & $0.483$   & $1.968$  & $10.643$           & $7.671$           & $2520.6$  \\
MASTER OT J084404.01+794408.7 & 1144277041110541440 & 1144277041110539136 & $1.878 \pm 0.096$    & $2.334 \pm 0.296$   & $0.427$   & $2.680$  & $9.470$            & $11.337$          & $12765.8$ \\
BP CrA                        & 6733050836430023552 & 6733049290241787392 & $1.780 \pm 0.042$    & $1.719 \pm 0.017$   & $0.492$   & $0.736$  & $6.055$            & $4.477$           & $27450.2$ \\
RZ Gru                        & 6544371342567818496 & 6544371346862823424 & $1.777 \pm 0.020$    & $1.735 \pm 0.080$   & $0.204$   & --       & $3.641$            & $7.649$           & $1049.0$  \\
VZ Scl                        & 2337436792938619392 & 2337437617572340480 & $1.735 \pm 0.041$    & $1.770 \pm 0.019$   & $0.382$   & $0.871$  & $6.578$            & $5.246$           & $32577.8$ \\
ASASSN-14eq                   & 4918835764173746432 & 4918835798533484288 & $1.586 \pm 0.076$    & $1.557 \pm 0.034$   & $0.644$   & $1.364$  & $8.770$            & $6.949$           & $27415.9$ \\
NSV 7184                      & 5985406470331309312 & 5985406843929482752 & $1.461 \pm 0.043$    & $1.386 \pm 0.260$   & $0.802$   & $2.470$  & $7.058$            & $9.438$           & $11513.1$ \\
ASASSN-18ed                   & 6139397368696634240 & 6139397368696633600 & $1.381 \pm 0.142$    & $1.674 \pm 0.017$   & $0.373$   & $0.707$  & $9.407$            & $4.725$           & $9604.1$  \\
AY Psc                        & 2565601982736199168 & 2565601982736199296 & $1.368 \pm 0.043$    & $1.388 \pm 0.088$   & $0.812$   & $1.768$  & $6.778$            & $7.864$           & $28801.9$ \\
SDSS J080033.86+192416.5      & 670132550216853632  & 670132545920724224  & $1.335 \pm 0.245$    & $1.032 \pm 0.190$   & $0.571$   & $2.059$  & $10.273$           & $8.636$           & $3334.6$  \\
ASASSN-19wi                   & 1777641510176281600 & 1777640754262037248 & $1.333 \pm 0.058$    & $1.373 \pm 0.183$   & $1.521$   & $0.533$  & $7.080$            & $9.346$           & $2064.5$  \\
ASASSN-14jd                   & 5269753451459457024 & 5269753455752368768 & $1.257 \pm 0.110$    & $1.340 \pm 0.081$   & $0.202$   & $1.226$  & $7.726$            & $6.917$           & $936.1$   \\
ASASSN-19eu                   & 6099000383784197120 & 6099000379484471168 & $1.176 \pm 0.090$    & $1.454 \pm 0.019$   & $0.675$   & $0.513$  & $6.051$            & $2.730$           & $1404.2$  \\
SWIFT J2124.6+0500            & 2699191408560964736 & 2699191404267020544 & $1.136 \pm 0.022$    & $1.108 \pm 0.077$   & $0.194$   & $1.329$  & $2.383$            & $6.681$           & $5279.2$  \\
TX Col                        & 4804695423438691200 & 4804695427734393472 & $1.020 \pm 0.020$    & $1.065 \pm 0.026$   & $0.255$   & $1.148$  & $5.079$            & $5.656$           & $2525.0$ 
\enddata
\end{deluxetable}
\end{longrotatetable}

\FloatBarrier
\bibliography{references}
\end{document}